\documentclass[%
preprint,
 amsmath,amssymb,
 aps, physrev,
]{revtex4-2}
\usepackage{natbib}

\usepackage{graphicx}
\usepackage[normalem]{ulem}
\usepackage{dcolumn}
\usepackage{bm}
\usepackage{xcolor}
\usepackage{comment}

\newcommand{\beq}{\begin{equation}}
\newcommand{\eeq}{\end{equation}}

\begin{document}
\raggedbottom 

\title{Mitigation of Initial Transients in Total-f Gyrokinetic Turbulence Simulations Using Neoclassically Relaxed Distribution Function}
\author{S. Ku}
\email{Contact author: sku@pppl.gov}
\author{R. Hager}
\author{C.S. Chang} \affiliation{
 Princeton Plasma Physics Laboratory, Princeton, NJ, 08540, USA}
\author{S.-J. Lee}
\affiliation{ITER Organization, St. Paul Lez Durance 13067, France}
\author{A. Scheinberg}
\affiliation{Jubilee Development, Cambridge, MA, United States}

\date{\today}
\begin{abstract}

Total-f five-dimensional gyrokinetic simulations are essential for self-consistent studies of multiscale, multiphysics transport in the edge region of diverted tokamak plasmas. However, conventional initialization with a local Maxwellian distribution often generates large-amplitude transients, particularly geodesic acoustic modes (GAMs). These transients are especially severe in the plasma edge because of steep profile gradients, strong radial electric fields, and high safety factors, and they can interfere with early-time turbulence and transport diagnostics. To address this problem, we present a new initialization scheme for the total-f XGC code that uses a relaxed particle distribution obtained from a computationally inexpensive axisymmetric simulation. Before the distribution is transferred to the full turbulence simulation, phase-space smoothing is applied to reduce particle noise while preserving its neoclassical structure. Simulations of the Cyclone Base Case and an ASDEX Upgrade I-mode discharge show that the method reduces particle noise and substantially suppresses initialization-driven transients.
\end{abstract}

\maketitle

\section{\label{sec:introduction} Introduction}

%

Gyrokinetic simulations are important tools for predicting turbulent transport in magnetically confined fusion plasmas. In the plasma edge, including the pedestal and scrape-off layer (SOL), a total-f treatment is especially important because the turbulence dynamics, the neoclassical dynamics, and the sharp plasma profile evolution occur on comparable spatial and temporal scales. In this regime, steep density, temperature, and electric-field gradients make finite-orbit-width effects and magnetic-drift transport inseparably interact together, so the total distribution function must be evolved self-consistently. This motivates the use of total-f/full-f gyrokinetic codes such as XGC\cite{Ku_2018}, GENE-X\cite{MICHELS_2021}, Gkeyll\cite{Shi_2017}, and GYSELA\cite{Grandgirard_2016} for edge transport studies. 

Total-f gyrokinetic codes solve an initial-value problem and therefore require an initial particle distribution function. A common choice is a local Maxwellian constructed from flux-surface-averaged profiles of density, temperature, and toroidal rotation. This distribution subsequently relaxes toward a non-Maxwellian state consistent with the plasma equilibrium. Although convenient, the initial Maxwellian can differ substantially from the equilibrium distribution.

This mismatch can drive a rapid relaxation of the system and excite large-amplitude, long-wavelength axisymmetric transients, most notably the Geodesic Acoustic Mode (GAM). These transients are especially problematic in the edge. 
Steep profile gradients and strong radial electric fields ($E_r$) modify particle trajectories and the equilibrium distribution, increasing the mismatch with the initial local Maxwellian. In addition, the safety factor ($q$) is typically large in the edge, which reduces the collisionless Landau damping rate of GAMs\cite{Winsor_1968, Xu_2009}. Combined with the low GAM frequencies ($\omega_{GAM} \propto v_{Ti}/R$), this effect can produce oscillations that persist for hundreds of microseconds and overlap with the early development of turbulence\cite{Grandgirard_2007}.

A canonical Maxwellian can improve upon local-Maxwellian initialization\cite{Idomura_2003, Angelino_2006, Lanti_2020} by incorporating conserved quantities such as the canonical toroidal momentum ($P_\phi$) and thus capturing finite-orbit-width effects more faithfully. However, constructing a canonical Maxwellian remains difficult in the steep-gradient edge, where density, temperature, toroidal angular momentum, and electric-field profiles can all vary sharply.
The limitations of the canonical-Maxwellian initialization are discussed in Sec.~\ref{sec:transient}.

We propose a neoclassical initialization scheme in the multiscale, multiphysics gyrokinetic code XGC that uses a numerically relaxed non-Maxwellian distribution function obtained from an inexpensive axisymmetric neoclassical simulation. By applying phase-space smoothing to the relaxed axisymmetric distribution function before it is loaded into the total-f calculation, this method provides a substantially more quiescent and physically accurate starting point than the conventional analytic Maxwellian-based initialization. We implement and test this approach in XGC. The strategy of starting from a numerically relaxed axisymmetric state, together with its self-consistent $E_r$, can also be used in other total-f/full-f codes, including grid-based ones; the smoothing operators, in contrast, address marker noise and are specific to particle-in-cell codes.


The remainder of this paper is organized as follows. In Sec.~\ref{sec:transient}, we describe the origin and properties of the initial transient oscillations and how a canonical Maxwellian particle distribution can reduce the transient. In Sec.~\ref{sec:scheme}, we present the axisymmetric numerical initialization scheme. In Sec.~\ref{sec:cbc}, we demonstrate its effectiveness in Cyclone Base Case (CBC) geometry. In Sec.~\ref{sec:asdex}, we apply the method to a realistic ASDEX Upgrade I-mode geometry. Finally, in Sec.~\ref{sec:conclusion}, we summarize the main findings and conclusions.

\section{Initial transient oscillations}\label{sec:transient}

\subsection{Gyrokinetic equations}
In this study, we use XGC, a five-dimensional gyrokinetic total-f particle-in-cell (PIC) code capable of treating neoclassical transport, turbulence, and neutral-particle dynamics together. XGC is designed for edge and scrape-off-layer physics, but its simulation domain can extend from the magnetic axis to the material wall.

XGC uses the hybrid-Lagrangian total-f scheme described in Sec.~II~B of Ku \textit{et al.}\cite{Ku_2018}. In the notation used here, the distribution is decomposed as $f=f_M+f_{\rm grid}+f_{\rm ptl}$. Here, $f_M$ is an analytic local Maxwellian, $f_{\rm grid}$ is a correction stored on the five-dimensional phase-space mesh, and $f_{\rm ptl}$ is the remaining correction represented by weighted marker particles. The combined background $f_M+f_{\rm grid}$ can therefore be non-Maxwellian and can vary in time. The scheme evolves the total distribution, retaining the magnetic-drift contribution to its evolution without assuming that $f$ remains close to a Maxwellian. A fraction of the marker contribution can be transferred to $f_{\rm grid}$ to reduce marker-weight growth while preserving the total distribution.

XGC solves the 5D gyrokinetic Boltzmann equation,
\beq
\partial_t f_s +
\dot{\mathbf{X}} \cdot \nabla f_s +
\dot{u}_\parallel \partial_{u_\parallel}\! f_s
= S(f_s)
\label{eq:boltzmann}
\eeq
where $f_s$ is the distribution of plasma species $s$ in 5D gyrokinetic phase space $f_s(\mathbf{X}, u_\parallel, \mu)$, 
$\mathbf{X}$ is the gyrocenter in the configuration space,  
$u_{\parallel}$ is the parallel velocity coordinate,
$\mu$ is the magnetic moment, 
and $S(f)$ is the phase-space operator such as Coulomb collisions. In general, $S(f)$ is a phase space operator that can break the phase-space volume conservation and change entropy.
The time derivatives of $\mathbf{X}$ and $u_\parallel$ follow from the Lagrangian equations of motion\cite{Littlejohn_1985,Hahm1988, Hager_2022}. For electromagnetic perturbations, XGC uses the mixed-variable formulation\cite{Mishchenko_2014}, as given in Eqs.~(5)--(8) of Hager \textit{et al.}\cite{Hager_2022}.
In this study, we restrict the analysis to the electrostatic perturbations for simplicity, for which the phase-space equations of motion become  

\begin{align}
\dot{\mathbf{X}} &=
u_{\parallel}\frac{\mathbf{B}^{*}}{B_{\parallel}^{*}}
+\frac{\hat{\mathbf{b}}}{q_s B_{\parallel}^{*}}
\times
\left(
\mu \nabla  B + q_s \overline{ \nabla \phi}
\right),
\\
\dot{u}_{\parallel} &=
-\frac{\mathbf{B}^{*}}{m_s B_{\parallel}^{*}}
\cdot
\left(
\mu\nabla B
+q_s \overline{\nabla \phi} 
\right),
\\
\mathbf{B}^{*} &= \mathbf{B} + \frac{m_s}{q_s} u_{\parallel}\,\nabla \times \hat{\mathbf{b}},
\\
B_{\parallel}^{*} &= \mathbf{B}^{*} \cdot \hat{\mathbf{b}}
\\
u_{\parallel} &= \hat{\mathbf{b}} \cdot \dot{\mathbf{X}},
\end{align}
where $\mathbf{B}$ is the magnetic field, 
$\hat{\mathbf{b}}=\mathbf{B}/B$, 
$B=|\mathbf{B}|$,  
$\phi$ is the electrostatic potential,
$q_s$ and $m_s$ are the charge and the mass of species $s$ respectively,
and $\overline{Y}$ represents gyro-average operation of a field quantity $Y$.
The electrostatic potential $\phi$ is determined from the gyrokinetic Poisson equation\cite{Hager_2022}, which in the Pad\'{e} approximation reads
\beq
\nabla\cdot\frac{n_0 m_i}{q_i B_0^2}\left(1+\frac{T_i}{T_e}\right)\nabla_\perp\phi = -\left(1-\nabla\cdot\rho_i^2\nabla_\perp\right)\left(\overline{n}_i-n_e\right),
\label{eq:poisson}
\eeq
where $n_0$ is the background density, $T_i$ and $T_e$ are the ion and electron temperatures, $B_0$ is the equilibrium magnetic-field strength, $\overline{n}_i$ is the gyroaveraged ion gyrocenter density, $n_e$ is the electron density, and $\rho_i=(m_iT_i)^{1/2}/(q_iB_0)$ is the ion gyroradius. The term proportional to $T_i/T_e$ on the left-hand side is the Pad\'{e} finite-Larmor-radius correction to the adiabatic electron response, and the term proportional to $\rho_i^2$ on the right-hand side is the corresponding correction to the charge density. Both are of order $k_\perp^2\rho_i^2$ and vanish in the long-wavelength limit. They are retained in the non-axisymmetric ($n\neq0$) field solve of the three-dimensional turbulence calculations, whereas the axisymmetric ($n=0$) solve, including the axisymmetric precursor calculations, uses the long-wavelength form. When Coulomb collisions are included, they are evaluated with the fully nonlinear multi-species Fokker--Planck--Landau operator of Hager \textit{et al.}\cite{Hager_2016} on the velocity mesh described in Sec.~\ref{sec:representation}. Further details of the field solve and the collision operator in the total-f scheme are given in Ref.~\onlinecite{Ku_2018} and the references therein.

\subsection{Phase-space representation and meshes}\label{sec:representation}

Each species is described in the five-dimensional gyrocenter phase space $\mathbf{z}=(\mathbf{X},u_\parallel,\mu)$, where $\mathbf{X}=(R,\varphi,Z)$ is the gyrocenter position in cylindrical coordinates, $u_\parallel$ is the parallel velocity, and $\mu$ is the magnetic moment; equivalently, $v_\perp=(2\mu B/m_s)^{1/2}$ is used as the perpendicular velocity coordinate.

\paragraph{Full-f, delta-f and total-f.}
In a full-f particle-in-cell method, marker particles sample the entire distribution, $f(\mathbf{z})\simeq\sum_j w_{0,j}\,\delta(\mathbf{z}-\mathbf{z}_j)$, where $w_{0,j}$ is the number of physical particles represented by marker $j$, and $f$ is evolved only through the marker motion. In a delta-f method, the distribution is split as $f=f_0+\delta f$ with a prescribed background $f_0$, usually a Maxwellian or a canonical Maxwellian, that is not evolved, and the markers represent only $\delta f$. 
Terms nonlinear in $\delta f$ are often neglected; in addition, the $\nabla B$ drift is often omitted from the weight evolution equation when the background is a local Maxwellian, and collisional terms from the background are often neglected when the background is a canonical Maxwellian. The total-f method used in XGC keeps the split $f=f_0+\delta f$ for noise reduction but evolves the full nonlinear kinetic equation, Eq.~(\ref{eq:boltzmann}), for the total $f$, and allows the background to evolve. The background is
\beq
f_0 = f_M + f_{\rm grid},
\eeq
where $f_M$ is an analytic local Maxwellian and $f_{\rm grid}$ is a non-Maxwellian correction stored on the phase-space mesh described below. The markers carry, in addition to $w_{0,j}$, a dimensionless weight $w_{1,j}$, and represent the remainder
\beq
\delta f = f - f_0 \simeq f_{\rm ptl} = \sum_j w_{1,j}\,w_{0,j}\,\delta(\mathbf{z}-\mathbf{z}_j).
\eeq
We write $\delta f=f-f_0$ for this marker part and reserve $\Delta f=f-f_M$ for the deviation from the local Maxwellian.
In the absence of collisions and sources, $w_{0,j}$ is constant along the marker trajectory and $w_{1,j}$ evolves so that the total $f$ is conserved along it, $dw_{1,j}/dt=-(df_0/dt)/(w_0 g)$ evaluated at $\mathbf{z}_j$ \cite{Ku_2018}, where g is the marker particle distribution. Part of $f_{\rm ptl}$ can be transferred to $f_{\rm grid}$ during the simulation, which keeps $|w_1|$ small without changing $f$. The size of $w_1w_0$ therefore measures the correction carried by the markers. This correction contains sampling noise as well as physical changes of the distribution relative to $f_0$, and its root-mean-square (rms) value [Eq.~(\ref{eq:wrms})] is used here as a proxy for particle noise.

\paragraph{Meshes.}
Two meshes are used.
(i) The configuration-space mesh is an unstructured triangular mesh in the poloidal $(R,Z)$ plane whose nodes are approximately aligned with the flux surfaces. The electrostatic potential, the field quantities, and $f_{\rm grid}$ are all defined at its nodes. The axisymmetric precursor uses a single poloidal plane and evolves only the $n=0$ dynamics, so the gyrokinetic Poisson equation is solved in two dimensions on a mesh that needs to resolve only neoclassical scales. The turbulence simulation uses $N_\varphi$ poloidal planes (16 for the ASDEX Upgrade case), with field-line-following interpolation between them, and a finer mesh whose radial resolution is of order $\rho_i$ (268,459 nodes per plane for ASDEX Upgrade).
(ii) At every configuration-space node, a velocity mesh that is uniform in $(u_\parallel,v_\perp)$, normalized to the local thermal speed $v_{\rm th}=(T/m_s)^{1/2}$, covers $|u_\parallel|\le v_{\max}$ and $0\le v_\perp\le v_{\max}$. For the CBC precursor, $v_{\max}=5v_{\rm th}$ with $41\times41$ points; the CBC turbulence calculation covers velocity space only up to $4v_{\rm th}$.
The product of the two meshes is the phase-space mesh on which $f_{\rm grid}$ is stored, collisions are evaluated, and the distribution is written for diagnostics.

\paragraph{Loading a non-Maxwellian initial distribution.}
When the simulation is started from a given distribution $f_{\rm init}$ on the mesh, the moments of $f_{\rm init}$ define $f_M$, and $f_{\rm grid}=f_{\rm init}-f_M$ is loaded directly onto the mesh. Markers are sampled from a marker distribution $g$, which in XGC can be an arbitrary function and is usually chosen to be flat in configuration space and in velocity space within the configuration and velocity space boundary. 
Each full-f weight is set to $w_{0,j}=f_0(\mathbf{z}_j)/g(\mathbf{z}_j)$, so that the markers reproduce $f_{\rm init}$. The weights $w_{1,j}$ are set to zero apart from a small random perturbation that seeds the turbulence. The entire non-Maxwellian structure of the initial state is therefore carried by $f_{\rm grid}$, and $\delta f$ starts from the noise level; $w_1$ grows only as $f$ departs from $f_0$. With a Maxwellian start, $f_{\rm grid}=0$, and the markers must build up the whole neoclassical deviation through $w_1$, which is the origin of the larger marker weights in Fig.~\ref{fig:cbc_noise}.

\subsection{Initial transient oscillations from local Maxwellian}

The usual choice of initial distribution in gyrokinetic simulation is a local Maxwellian with density and temperature taken to be flux functions. However, this initial state is generally not a kinetic equilibrium of a toroidal plasma, and the mismatch generates a large transient response at the beginning of the simulation.

To estimate the initial transient in an axisymmetric toroidal equilibrium, we assume that the background magnetic field is axisymmetric and that, during the earliest phase of the relaxation, the dominant perturbed field is the flux-surface-averaged electrostatic potential $\left< \overline{\phi} \right>$. If we substitute the local Maxwellian
$
f_M = n(\psi) \left( \frac{m_s}{2 \pi T(\psi)} \right)^{3/2}
\exp{\left( -\frac{E}{T(\psi)}\right)}
$
into the gyrokinetic Boltzmann equation, Eq.~(\ref{eq:boltzmann}), neglect the source term, and use the toroidal and poloidal symmetry of $f_M$, we obtain
\beq
\partial_t f_s  = - \dot{\psi} f_M \left( \frac{\partial_\psi n}{n} + \left(\frac{E}{T} - \frac{3}{2}\right) \frac{\partial_\psi T}{T} + q_s \frac{\partial_\psi \left< \overline{\phi} \right>}{T} \right),
\label{eq:dfdt}
\eeq
where $\psi$ is the poloidal magnetic flux divided by $2\pi$, $\dot{\psi} = \dot{\mathbf{X}} \cdot \nabla \psi$, and $E=\frac{1}{2} m_s u_\parallel^2 + \mu B$. Equation (\ref{eq:dfdt}) shows that, for general profiles, a local Maxwellian remains stationary only in the asymptotic limit of vanishing radial orbit excursion, that is, $\dot\psi=0$, in which case Eq.~(\ref{eq:dfdt}) is satisfied for arbitrary profiles. In a toroidal plasma, however, guiding centers sample a finite radial interval $\Delta \psi_{\rm orb}$ over one bounce or transit period, so the local Maxwellian is immediately distorted by the orbit excursion across the background gradients.

An order-of-magnitude estimate of the initial mismatch can be obtained by integrating Eq.~(\ref{eq:dfdt}) over one orbit period. We use the minor radial coordinate $r$ instead of $\psi$, with $\Delta r \sim \Delta \psi ~\partial_\psi r$. The radial orbit excursion is then $\Delta r_{\rm orb} \sim \tau_{\rm orb}\dot{\psi} \partial_\psi r$. With these definitions, the mismatch scales as
\beq
\frac{\Delta f_{\rm init}}{f_M}
\sim
\frac{\Delta r_{\rm orb}}{L_n} + 
\left(\frac{E}{T}-\frac{3}{2}\right)\frac{\Delta r_{\rm orb}}{L_T} +
\frac{e\,\delta \phi_{\rm init}}{T_i},
\label{eq:deltaf_init_r}
\eeq

where
\beq
L_n^{-1} \equiv \left| \partial_r \ln n \right|,
\qquad
L_T^{-1} \equiv \left| \partial_r \ln T \right|,
\eeq
up to geometry-dependent coefficients of order unity.

The corresponding initial ion-density perturbation is obtained by taking the velocity moment of $\Delta f_{\rm init}$. For a local Maxwellian, the temperature-gradient factor satisfies $\int d^3v\, f_M(E/T-3/2)=0$. During relaxation, however, finite orbit excursions generate a non-Maxwellian correction $\Delta f_{\rm init}$. Because the orbit displacement depends on particle energy and pitch angle, the temperature-gradient contribution to this correction carries an additional velocity-dependent weighting and need not cancel upon velocity integration. Both density and temperature gradients can therefore contribute to the density perturbation. For an order-of-magnitude estimate,
\beq
\frac{\delta n_i}{n_i}
\sim
\mathcal{O}\!\left(\frac{\Delta r_{\rm orb}}{L_n}\right)
+
\mathcal{O}\!\left(\frac{\Delta r_{\rm orb}}{L_T}\right).
\label{eq:density_imbalance_est}
\eeq
This initial charge imbalance drives the axisymmetric electrostatic response, and the amplitude of the resulting GAM increases with it.
This shows why the transient is weak in the core but becomes large in the edge pedestal region: there the orbit width $\Delta r_{\rm orb}$ becomes a non-negligible fraction of both $L_n$ and $L_T$.

This relaxation produces a rapid redistribution of charge and parallel flows, which excites axisymmetric zonal responses. The dominant component of this response is typically the Geodesic Acoustic Mode (GAM), an oscillatory $E_r$ perturbation coupled to compressional motion through magnetic geometry\cite{Conway_2022}. Since the perturbation is generated by the initialization itself rather than by turbulence or external forcing, the resulting GAM should be regarded as an artificial transient.

\subsection{Canonical Maxwellian}

A natural improvement over the local Maxwellian is the canonical Maxwellian, which incorporates constants of motion of an axisymmetric system into the equilibrium distribution \cite{Idomura_2003, Angelino_2006, Lanti_2020, McMillan_2011}. Instead of constructing the distribution solely from local fluid moments at a given position, this approach expresses the equilibrium in terms of orbit-related quantities, most notably the toroidal canonical momentum $P_\phi$, together with energy and magnetic moment. Because $P_\phi$ is conserved in an axisymmetric equilibrium, the resulting distribution is more consistent with the underlying particle trajectories than a purely local Maxwellian.

In gyrokinetic formalism, the toroidal canonical momentum is
\beq
P_\phi = m_s R \frac{B_\phi}{B} u_\parallel + q_s \psi,
\eeq
and we define $\psi_c = P_\phi / q_s$. In the canonical Maxwellian, density and temperature are taken to be functions of $\psi_c$, so the distribution becomes
\beq
f_{CM} = n(\psi_c) \left( \frac{m_s}{2 \pi T(\psi_c)} \right)^{3/2} \exp{\left( -\frac{E}{  T(\psi_c)}\right)}
\label{eq:cm}
\eeq
There is also a modified version of the canonical Maxwellian\cite{Angelino_2006} that uses
$$
\psi_{\rm corr} \equiv  \psi_c - {\rm sign}(u_\parallel) \frac{m_s}{q_s} R_0 \sqrt{\frac{2(E-\mu B_0)}{m_s}} ~  \mathcal{H}(E-\mu B_0),
$$
where $R_0$ and $B_0$ are the major radius and the magnetic-field strength at the magnetic axis, and $\mathcal{H}$ is the Heaviside function. This correction vanishes for trapped particles and has opposite signs for forward and backward passing particles. It reduces the passing-particle offset in $\psi_c$ and can mitigate the unrealistic parallel flows produced by Eq.~(\ref{eq:cm}).

The key advantage of the canonical Maxwellian is that it is written in terms of orbit-related invariants. In a static, axisymmetric, collisionless equilibrium with no electric field, a distribution expressed as a function of $\psi_c$, $E$, and $\mu$ is constant along particle trajectories. This can be seen from the counterpart of Eq.~(\ref{eq:dfdt}). Substituting $f_{CM}$ into Eq.~(\ref{eq:boltzmann}) without the collision and source terms gives
\beq
\partial_t f_{CM} = - \dot{\psi}_c f_{CM} \left( \frac{\partial_{\psi_c} n}{n} + \left(\frac{E}{T} - \frac{3}{2}\right) \frac{\partial_{\psi_c} T}{T} \right),
\label{eq:dfdt_cm}
\eeq
in which $\dot\psi_c=d\psi_c/dt$ along the gyrocenter trajectory replaces $\dot\psi$ of Eq.~(\ref{eq:dfdt}). When the magnetic field is axisymmetric ($n=0$) and does not change in time, and no turbulence breaks this symmetry, $P_\phi$, and hence $\psi_c$, is an invariant of the motion. Then $\dot\psi_c=0$ and the right-hand side of Eq.~(\ref{eq:dfdt_cm}) vanishes, whereas that of Eq.~(\ref{eq:dfdt}) vanishes, for general profiles, only in the limit of vanishing radial orbit excursion. This construction can therefore reduce the initial transient relative to a local Maxwellian defined only by flux-surface moments.

When a background radial electric field is present, the conserved energy in a static, axisymmetric electrostatic equilibrium is the total energy, $H=E+q_s\Phi(\psi)$, rather than the kinetic energy $E$ alone. McMillan \textit{et al.}\cite{McMillan_2011} addressed the inclusion of a background electrostatic potential by constructing a canonical distribution using the total energy together with a potential correction evaluated at an invariant radial coordinate. Thus, a spatially varying background potential can be incorporated while preserving the canonical structure of the distribution. Nevertheless, constructing a distribution that simultaneously reproduces the desired density, temperature, and parallel-flow profiles together with a self-consistent radial electric field remains difficult, particularly in the steep-gradient edge.

Moreover, determining the self-consistent $E_r$ for the target neoclassical state requires a neoclassical simulation, so a canonical-Maxwellian initialization based on that field entails a preparation simulation effort comparable to that of the method proposed here. Once this calculation has been performed, its relaxed distribution and self-consistent field can be used directly, avoiding the additional task of constructing a canonical Maxwellian that reproduces the same profiles and flows. This motivates our use of the numerical non-Maxwellian distribution obtained from the axisymmetric precursor calculation.

\section{Algorithm\label{sec:scheme}}

To reduce the initial transient, we initialize the full turbulence simulation from a numerically relaxed neoclassical state rather than from an analytic Maxwellian. The method uses both the distribution and the self-consistent radial electric field obtained from the axisymmetric precursor simulation as the initial condition for the production run. The workflow consists of three stages: (i) an inexpensive axisymmetric precursor calculation to obtain a neoclassical pseudo-equilibrium, including its self-consistent radial electric field, (ii) conditioning of the extracted 5D distribution to suppress particle noise while preserving the neoclassical-scale structure, and (iii) interpolation of the conditioned distribution and field quantities onto the target simulation mesh (neoclassical or for turbulence) for the production run.

\subsection{Axisymmetric precursor calculation}

The first stage is an axisymmetric ($n=0$) XGC simulation that is advanced until the initial transient oscillations have decayed. The purpose of this run is not to compute turbulence, but to establish a neoclassical pseudo-equilibrium that is substantially closer to the steady state of the subsequent total-f simulation than a local Maxwellian. In particular, the precursor simulation relaxes the ion distribution and the flux-surface-averaged electrostatic potential together, thereby determining the neoclassical radial electric field associated with the relaxed state. This self-consistent $E_r$ is retained when the conditioned distribution is transferred to the target simulation.

The precursor is XGC itself run in its axisymmetric mode, often referred to as XGCa \cite{XGC_web}: the same total-f marker algorithm, equations of motion and gyrokinetic Poisson equation are used, but on a single poloidal plane with only the $n=0$ potential, so the field solve is two-dimensional, and on a configuration-space mesh too coarse to resolve turbulence scales. The turbulence calculation uses the full three-dimensional mode of the same code, formerly called XGC1. For CBC, the precursor uses adiabatic electrons, no collisions, and about $9\times10^3$ ion markers per mesh node. For ASDEX Upgrade, it uses kinetic electrons, collisions, and about $10^4$ markers per mesh node for each species. The numerical parameters of the precursor and turbulence calculations are listed in Table~\ref{tab:runs}.

\begin{table}
\caption{Numerical parameters of the precursor and turbulence calculations. Markers are given per mesh node and per kinetic species. The velocity mesh is given as the number of points in $u_\parallel\times v_\perp$ and its extent $v_{\max}$. AUG denotes ASDEX Upgrade; the AUG turbulence calculation is the one started from the neoclassical initial state.}
\label{tab:runs}
\begin{ruledtabular}
\begin{tabular}{lccccc}
Calculation & $\Delta t$ (s) & Markers/node & Velocity mesh & Nodes/plane & Planes \\
\hline
CBC precursor, axisym.\ tests & $1.5\times10^{-7}$ & $9\times10^{3}$ & $41\times41$, $5v_{\rm th}$ & 8\,568 & 1 \\
CBC turbulence & $3.9\times10^{-7}$ & $9\times10^{3}$ & $31\times33$, $4v_{\rm th}$ & 107\,943 & 16 \\
AUG precursor & $7.6\times10^{-8}$ & $1\times10^{4}$ & $35\times34$, $4v_{\rm th}$ & 17\,550 & 1 \\
AUG turbulence & $1.5\times10^{-7}$ & $8\times10^{3}$ & $35\times34$, $4v_{\rm th}$ & 268\,459 & 16 \\
\end{tabular}
\end{ruledtabular}
\end{table}


This precursor step is computationally inexpensive compared with the full 3D turbulence simulation for two reasons. First, the toroidal non-axisymmetric modes are excluded, so only the $n=0$ dynamics are evolved. Second, the mesh in the configuration space can be significantly coarser than the turbulence mesh because it only needs to resolve neoclassical-scale structure rather than the turbulence correlation length, $ \mathcal{O}\!\left(\rho_i\right)$. In practice, the precursor calculation is more than two orders of magnitude cheaper than the turbulence run per unit of simulated time (about $400$ times for the ASDEX Upgrade case of Sec.~\ref{sec:asdex}) and adds a small preparation cost.

In the CBC case, Coulomb collisions are omitted for numerical efficiency from the precursor and from the subsequent relaxation and turbulence calculations. This follows the collisionless setting of the standard CBC benchmark\cite{Dimits_2000}; the relaxed distribution is then a collisionless pseudo-equilibrium rather than a collisional neoclassical equilibrium. In the ASDEX Upgrade case, collisions are included in both the precursor and the turbulence calculation, using the nonlinear Fokker--Planck--Landau operator.
The resulting distribution is therefore not claimed to be the exact neoclassical equilibrium, but rather a numerically relaxed pseudo-equilibrium that already captures the dominant orbit-width, magnetic-drift, and electrostatic effects responsible for the large artificial GAM transient. At the end of the axisymmetric run, the full five-dimensional distribution function is extracted on the XGC phase-space mesh.

\subsection{Distribution conditioning and smoothing}

\begin{figure}
\includegraphics[width=0.38\textwidth]{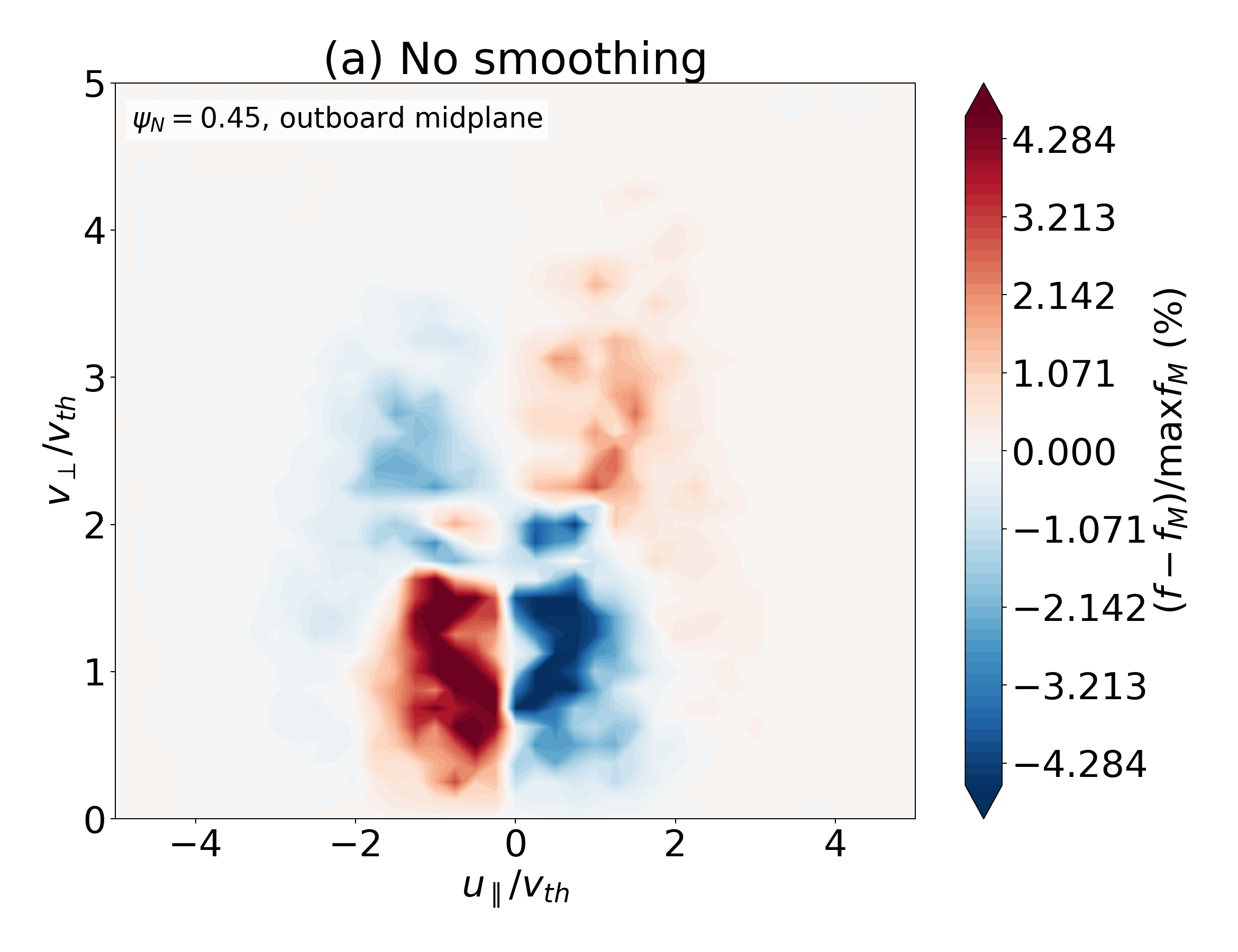}
\includegraphics[width=0.38\textwidth]{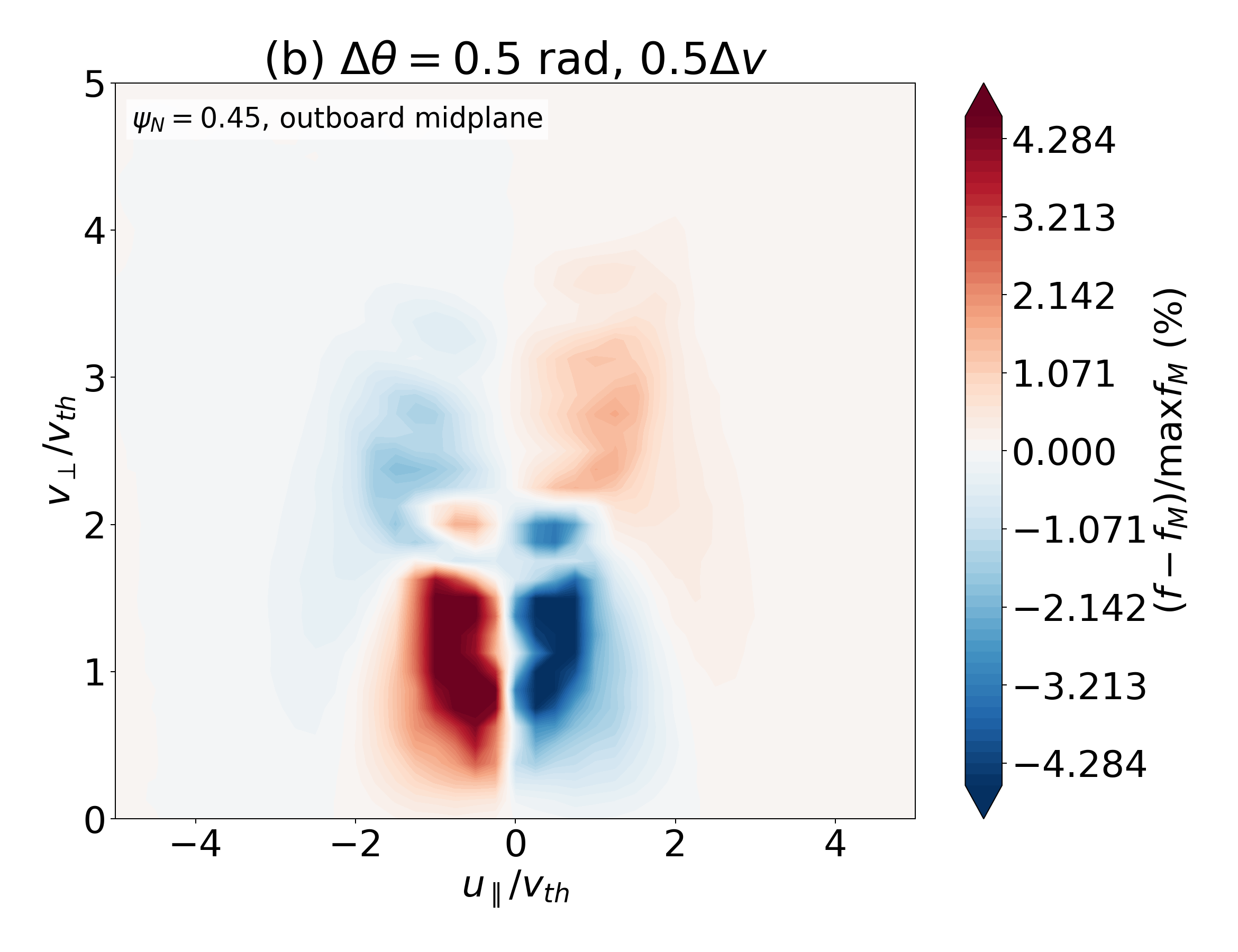}
\caption{
Deviation of the extracted distribution from the local Maxwellian, expressed as $100(f-f_M)/\max f_M$ (\%), where $\max f_M$ is the peak of the local Maxwellian over velocity space at the sampled node. The velocity coordinates $(u_\parallel,v_\perp)$ are normalized to the local ion thermal speed. The sample is taken at the outboard midplane, $\psi_N=0.45$, in the axisymmetric CBC precursor at $t=0.31$ ms. (a) Without smoothing. (b) After Gaussian smoothing in poloidal angle and velocity space, with widths of $0.5$ rad and $0.5\Delta v$, respectively. Both panels use the same diverging color scale centered at zero; the color range is clipped at the 98th percentile of panel (a).
} 

\label{fig:distribution}
\end{figure}

The extracted 5D distribution contains statistical noise because each phase-space cell is sampled by only a limited number of markers. Figure~\ref{fig:distribution}(a) shows its deviation from a local Maxwellian in the $(u_\parallel,v_\perp)$ plane at the outboard midplane, $\psi_N=0.45$, in the Cyclone Base Case (CBC) simulation described in Sec.~\ref{sec:cbc}.
In the present implementation, the phase-space mesh contains only $\mathcal{O}(10)$ particles per 5D phase-space cell, even though each configuration-space cell contains roughly $10^4$ particles and the velocity-space mesh is typically about $40 \times 40$. Directly loading this noisy distribution into the turbulence calculation would seed unnecessary particle-weight growth and could interfere with the early-time dynamics.

To reduce this noise, we apply smoothing operators that target the neoclassical equilibrium scale while avoiding distortion of the physically important large-scale structure. The first operation is poloidal smoothing, which filters variations much shorter than the broad $k_{\theta} \sim 1/r_0$ structure of the axisymmetric equilibrium, where $k_{\theta}$ is poloidal wavenumber and $r_0$ is the minor radius. The dominant neoclassical poloidal structure is expected to be the $m=1$ mode, whose poloidal wavelength is $2\pi$. The smoothing width $\Delta\theta=0.5$ rad, for example, therefore corresponds to only $\Delta\theta/(2\pi) \simeq 0.08$ of this dominant poloidal scale, so it is expected to reduce short-scale marker noise without strongly distorting the neoclassical structure. The smoothing is applied in the poloidal angle along each flux surface rather than in a field-aligned angle, because the axisymmetric precursor distribution is not field-aligned turbulence; its structure is set by the neoclassical orbit dynamics on the flux surface. The second operation is velocity-space smoothing using a Gaussian kernel, which removes fine-scale Monte Carlo noise in the $(u_\parallel,v_\perp)$ dependence of the distribution.

Both operations act on the non-Maxwellian part $\Delta f=f-f_M$, as stored on the XGC velocity mesh including the $v_\perp$ Jacobian, and use a normalized discrete Gaussian kernel,
\beq
\Delta f_i \rightarrow \sum_k w_k\,\Delta f_{i+k}, \qquad w_k=\frac{\exp\left(-k^2/2\sigma^2\right)}{\sum_{k'}\exp\left(-k'^2/2\sigma^2\right)},
\label{eq:gauss}
\eeq
where $i$ is a mesh index and the standard deviation $\sigma$ is measured in mesh cells. For poloidal smoothing, $i$ runs over the $N_s$ nodes of one flux surface ordered in poloidal angle, the sum is periodic on closed surfaces, and $\sigma=N_s\Delta\theta/2\pi$, so that $\Delta\theta$ is the standard deviation in poloidal angle for uniformly spaced nodes. For velocity-space smoothing, the kernel is applied at every configuration-space node along both directions of the uniform $(u_\parallel,v_\perp)$ mesh, whose spacings $\Delta u_\parallel$ and $\Delta v_\perp$ are fixed in units of the local thermal speed $v_{\rm th}=\sqrt{T/m_i}$, with the same $\sigma$ in mesh cells in both directions, truncation at $4\sigma$, and reflection at the mesh boundaries. A width of $0.5\Delta v$ denotes $\sigma=0.5$ cells; on the CBC mesh, with $\Delta u_\parallel=0.25v_{\rm th}$ and $\Delta v_\perp=0.125v_{\rm th}$, this corresponds to standard deviations of $0.125v_{\rm th}$ in $u_\parallel$ and $0.0625v_{\rm th}$ in $v_\perp$.

After smoothing, a density correction is applied to reduce the flux-surface-averaged mismatch between the initial ion gyrocenter density and that of the precursor. XGC is first run from the conditioned distribution with $\Delta t=0$, so that the markers do not move and the ion gyrocenter density $n_0(\mathbf{x})$ reconstructed from the loaded marker and mesh contributions is the one implied by the initial distribution alone. With $n_{\rm ref}(\mathbf{x})$ the corresponding density of the precursor at the extraction time, the distribution is rescaled node by node,
\beq
f(\mathbf{x},\mathbf{v}) \rightarrow \left[1+\frac{\langle n_{\rm ref}-n_0\rangle_\psi(\mathbf{x})}{n_f(\mathbf{x})}\right] f(\mathbf{x},\mathbf{v}),
\label{eq:dn_adjust}
\eeq
where $\langle\cdot\rangle_\psi$ is the flux-surface average and $n_f(\mathbf{x})=\int f\,d^3v$ is the local density of the distribution being corrected. The moments of $f_M$ are kept fixed, so the change enters $\Delta f$. For the CBC distribution smoothed with $\Delta\theta=0.5$ rad and $0.5\Delta v$, the flux-surface-averaged mismatch before correction was $3\times10^{-5}$ of the local density in rms ($1.7\times10^{-4}$ at most), and a single pass was applied.

\subsection{Transfer to the turbulence calculation}

The conditioned distribution is then interpolated from the precursor mesh onto the finer turbulence mesh, whose radial resolution is of order $\rho_i$. In the XGC total-f formulation, this processed non-Maxwellian distribution is used consistently in both parts of the background representation: it is used to initialize the marker particles and to populate the 5D mesh component of the total distribution. The turbulence simulation is then launched from this neoclassical initial state without an additional relaxation stage.

This procedure has two numerical benefits. First, because the initial state is already close to the neoclassical pseudo-equilibrium, the artificial GAM response is strongly suppressed. Second, the required particle-weight correction is reduced, which lowers the variance of marker weights and therefore reduces particle noise during the early phase of the turbulence simulation. The benefits are demonstrated in the following sections.

\section{Cyclone Base Case Demonstration\label{sec:cbc}}

As a first validation of the initialization scheme, we consider the Cyclone Base Case (CBC), which provides a convenient core-plasma test problem for isolating the effect of the initialization procedure. The equilibrium is defined locally with major radius $R_0=1.70$ m, minor radius $a=0.61$ m ($R_0/a=2.78$), magnetic field $B_0=1.91$ T on axis, and a quadratic safety-factor profile, $q\simeq0.85+2.2(r/a)^2$. At $r/a=0.5$ ($\psi_N\simeq0.39$), $q=1.41$, the magnetic shear is $\hat s=0.78$, $R_0/L_T=6.5$, $R_0/L_n=2.1$, $T_e/T_i=1$, $T_i=2.3$ keV, and $a/\rho_i\simeq170$ with $\rho_i=(m_iT_i)^{1/2}/(eB_0)$. These values are close to the standard CBC parameters\cite{Dimits_2000}. The ions are deuterium, and the electrons are adiabatic. 
The axisymmetric XGC precursor calculation starts from a local Maxwellian. The non-Maxwellian distribution is extracted at $t=0.31$ ms, after the dominant initial transient has decayed, and processed using the smoothing procedure described in Sec.~\ref{sec:scheme}. The conditioned distribution is used in the axisymmetric relaxation tests. The turbulence calculation starts from a distribution prepared in the same way from a precursor calculation whose velocity mesh extends to $4v_{\rm th}$, smoothed with $\Delta\theta=0.5$ rad and $0.5\Delta v$, and interpolated onto a mesh with a spatial resolution of the order of the ion Larmor radius $\rho_i$.

\begin{figure}
\includegraphics[width=0.58\textwidth]{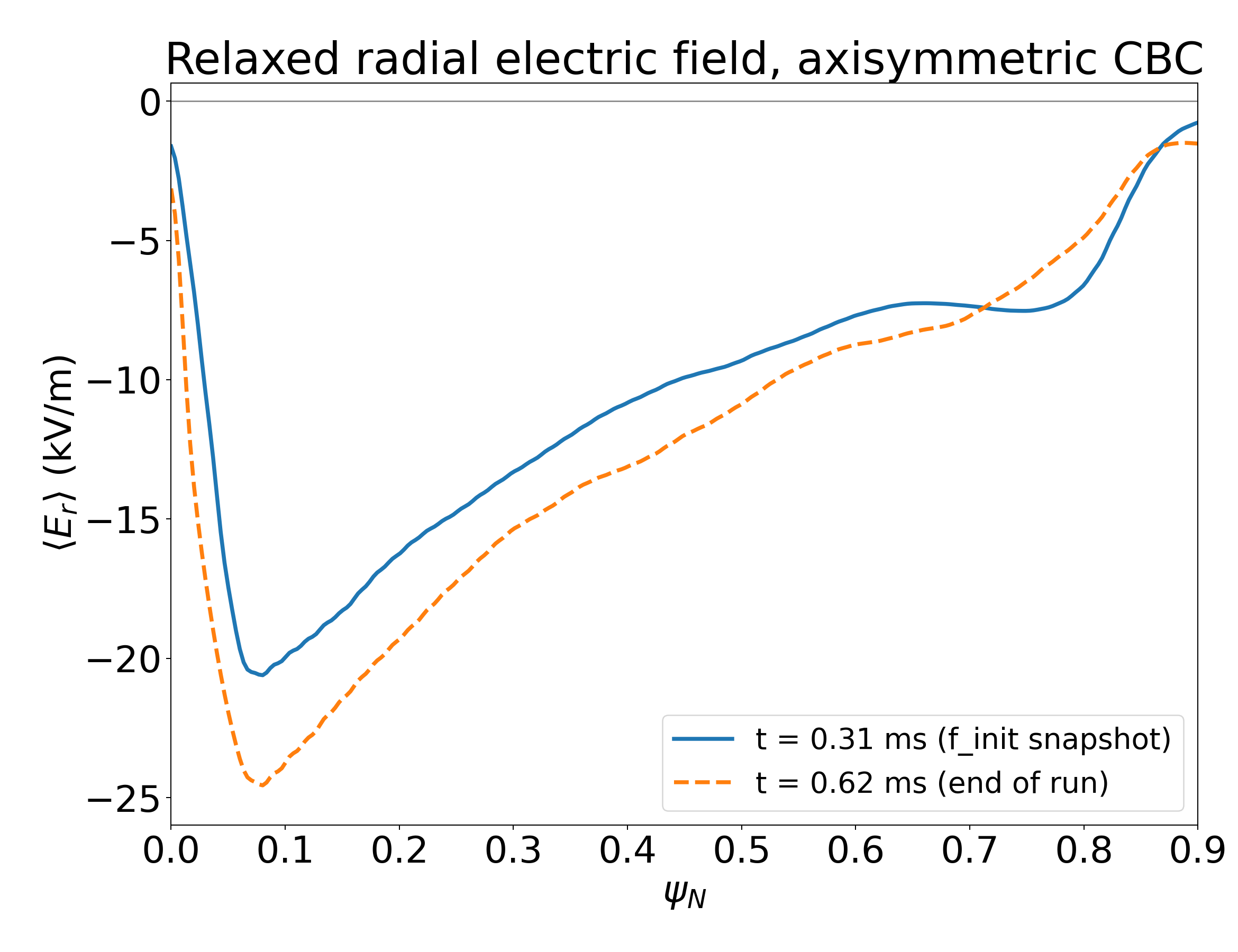}
\caption{Flux-surface-averaged radial electric field $\langle E_r \rangle = -(\partial \langle\phi\rangle/\partial\psi)\,\langle |\nabla\psi|^2 \rangle^{1/2}$, where $\langle\phi\rangle$ is the flux-surface-averaged electrostatic potential, in the axisymmetric CBC precursor at the distribution extraction time, $t=0.31$ ms, and at the end of the run, $t=0.62$ ms. The profiles are smoothed with a Gaussian of width $0.005$ in $\psi_N$, and the plotted interval ends at $\psi_N=0.9$ to exclude the outer boundary layer of the simulation domain.}
\label{fig:er_cbc}
\end{figure}

Figure~\ref{fig:er_cbc} shows the self-consistent radial electric field at the extraction time and at the end of the precursor run, $t=0.62$ ms. At extraction, $\langle E_r\rangle$ is negative throughout the plotted interval, reaching about $-20$ kV/m near $\psi_N\simeq0.1$ and decreasing in magnitude to a few kV/m toward the outer boundary. The later profile retains this overall shape but shows further evolution. The neoclassical initialization transfers this self-consistent field together with the relaxed distribution and its associated poloidal and parallel flows.

The CBC case is useful for two reasons. First, it allows a controlled comparison of the early-time transient response between Maxwellian and neoclassical initialization without the additional complexity of diverted edge geometry. Second, it provides a clear measure of the numerical noise level through the particle-weight evolution in the XGC total-f formulation. In this section, we therefore focus on three quantities: the residual transient amplitude in the axisymmetric test, the particle-weight level after initialization, and the early-time behavior of the turbulence simulation.

The transport and flow diagnostics are taken from the flux-surface-averaged one-dimensional output of XGC for the ion species. With $\langle\cdot\rangle$ the flux-surface average and $V(\psi)$ the volume enclosed by the flux surface $\psi$, the radial ion gyrocenter particle flux and energy flux through a flux surface are
\begin{align}
\Gamma_i(\psi) &= \frac{dV}{d\psi}\left\langle \int f_i\, \dot{\mathbf{X}}\cdot\nabla\psi \, d^3v \right\rangle,
\label{eq:pflux}\\
Q_i(\psi) &= \frac{dV}{d\psi}\left\langle \int \frac{m_i v^2}{2}\, f_i\, \dot{\mathbf{X}}\cdot\nabla\psi \, d^3v \right\rangle,
\label{eq:eflux}
\end{align}
where $\dot{\mathbf{X}}$ is the gyrocenter velocity given above and $v^2=u_\parallel^2+v_\perp^2$. $\Gamma_i$ and $Q_i$ are the number of ions and the kinetic energy crossing the whole flux surface per unit time. The $E\times B$ part of the energy flux, $Q_i^{E}$, is obtained by retaining only the $E\times B$ drift, $\hat{\mathbf{b}}\times\overline{\nabla\phi}/B_\parallel^*$, in $\dot{\mathbf{X}}\cdot\nabla\psi$.
Because the axisymmetric response is set by the balance between the ion gyrocenter charge flux and the polarization charge, we express the ion particle flux as a radial vorticity flux,
\beq
\Pi_\varpi(\psi) = -\left\langle \frac{\Omega_i}{n_{i0}} \right\rangle \Gamma_i(\psi), \qquad \Omega_i=\frac{eB}{m_i},
\label{eq:vortflux}
\eeq
where $n_{i0}$ is the initial ion density and $m_i$ is the deuterium mass in both cases considered here. $\Pi_\varpi$ has units of m$^3$\,s$^{-2}$. The $E\times B$ shear is evaluated from the flux-surface-averaged poloidal component of the ion $E\times B$ drift, $\langle V_{E,\theta}\rangle$, as
\beq
\omega_E(\psi) = \frac{\partial \langle V_{E,\theta}\rangle}{\partial\psi}\,\langle|\nabla\psi|^2\rangle^{1/2},
\label{eq:shear}
\eeq
and is shown in units of $10^3$ s$^{-1}$ (kHz).
The two diagnostics show the GAM transient from complementary sides. The vorticity flux is proportional to the radial ion gyrocenter charge flux $e\Gamma_i$. In the CBC tests, where the electrons are adiabatic, a net ion charge flux across a flux surface must be balanced by a change of the polarization charge, that is, of the $E\times B$ vorticity inside that surface. $\Pi_\varpi$ therefore measures the charge transport that drives the axisymmetric $E\times B$ flow: it oscillates at the GAM frequency about the small neoclassical flux, and it is obtained directly from the flux without the radial derivative required for the shear. The $E\times B$ shear, in turn, is the quantity through which the GAM acts on turbulence, and it is used here to assess the physical consequence of the transient.

\subsection{Axisymmetric test and poloidal smoothing}

\begin{figure}
\includegraphics[width=0.9\textwidth]{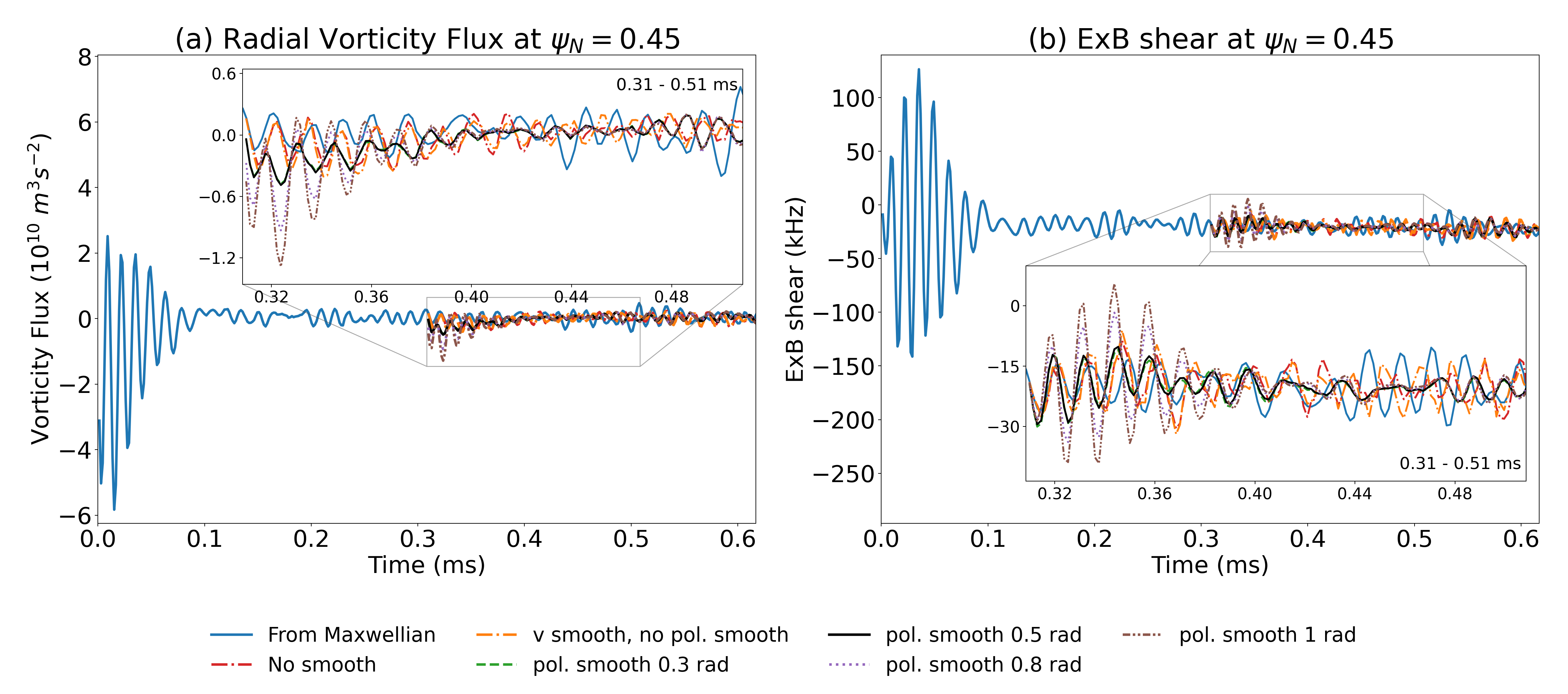}
\caption{Initial transient oscillations at $\psi_N=0.45$ in the axisymmetric CBC relaxation tests. (a) Radial vorticity flux through the flux surface, $\Pi_\varpi$ of Eq.~(\ref{eq:vortflux}), in units of $10^{10}$ m$^3$\,s$^{-2}$. (b) Flux-surface-averaged $E\times B$ shear, $\omega_E$ of Eq.~(\ref{eq:shear}). The Maxwellian run (solid blue) starts at $t=0$. The neoclassically initialized runs are offset to start at $t=0.31$ ms, the extraction time in the precursor run. These runs use the extracted distribution without smoothing, with velocity-space smoothing only ($0.5\Delta v$), or with velocity-space smoothing and poloidal smoothing widths of $\Delta\theta=0.3$, $0.5$ (solid black), $0.8$, and $1$ rad. The insets show $t=0.31$--$0.51$ ms, the first $0.2$ ms after neoclassical initialization.}
\label{fig:GAM}
\end{figure}

In an ideal implementation, initialization from the relaxed neoclassical distribution would produce little or no subsequent axisymmetric oscillation. In practice, however, the loaded state can differ slightly from the relaxed distribution because of numerical operations required during initialization. These include smoothing of the extracted 5D distribution, interpolation from the precursor mesh to the target mesh, and approximate treatment of the gyro-averaged ion density used to maintain charge consistency. These small inconsistencies can introduce a residual charge imbalance at the initial time and excite a low-amplitude GAM even in the neoclassical-start cases. The extent of the velocity mesh also matters: all axisymmetric CBC calculations shown here use a velocity mesh extending to $5v_{\rm th}$, and in our tests reducing it to $4v_{\rm th}$ increased the initial oscillation of the neoclassical starts. The CBC turbulence calculations were performed before this sensitivity was identified and use $4v_{\rm th}$; the neoclassical start nevertheless suppresses the GAM substantially (Sec.~\ref{sec:cbc_turb}).

Figure~\ref{fig:GAM} compares the initial axisymmetric relaxation for the different initialization procedures. Maxwellian initialization produces the largest oscillations in both diagnostics. Their frequency agrees with the analytic GAM frequency across the radial domain, which identifies them as GAMs; the frequency analysis is described in Appendix~\ref{app:gam}. All neoclassical initializations substantially reduce this response: the peak absolute vorticity flux decreases from about $6$ to $0.3$--$1.3$ in the units of the figure, while the peak absolute $E\times B$ shear decreases from about $140$ kHz to $30$--$40$ kHz. The magnitude of the reduction varies with the diagnostic and smoothing width. The unsmoothed distribution, velocity-space smoothing alone, and poloidal smoothing at $\Delta\theta=0.3$ and $0.5$ rad yield similar initial transients. Wider filters, $\Delta\theta=0.8$ and $1$ rad, produce larger initial vorticity-flux excursions. Thus, poloidal smoothing does not further reduce the initial transient in these tests. The larger response at wider filter widths is consistent with smoothing-induced distortion of the relaxed state, which is quantified in Appendix~\ref{app:distortion}. The benefit of poloidal smoothing is instead evident in the late-time residual oscillations and marker-weight levels discussed below.

\begin{figure}
\includegraphics[width=0.78\textwidth]{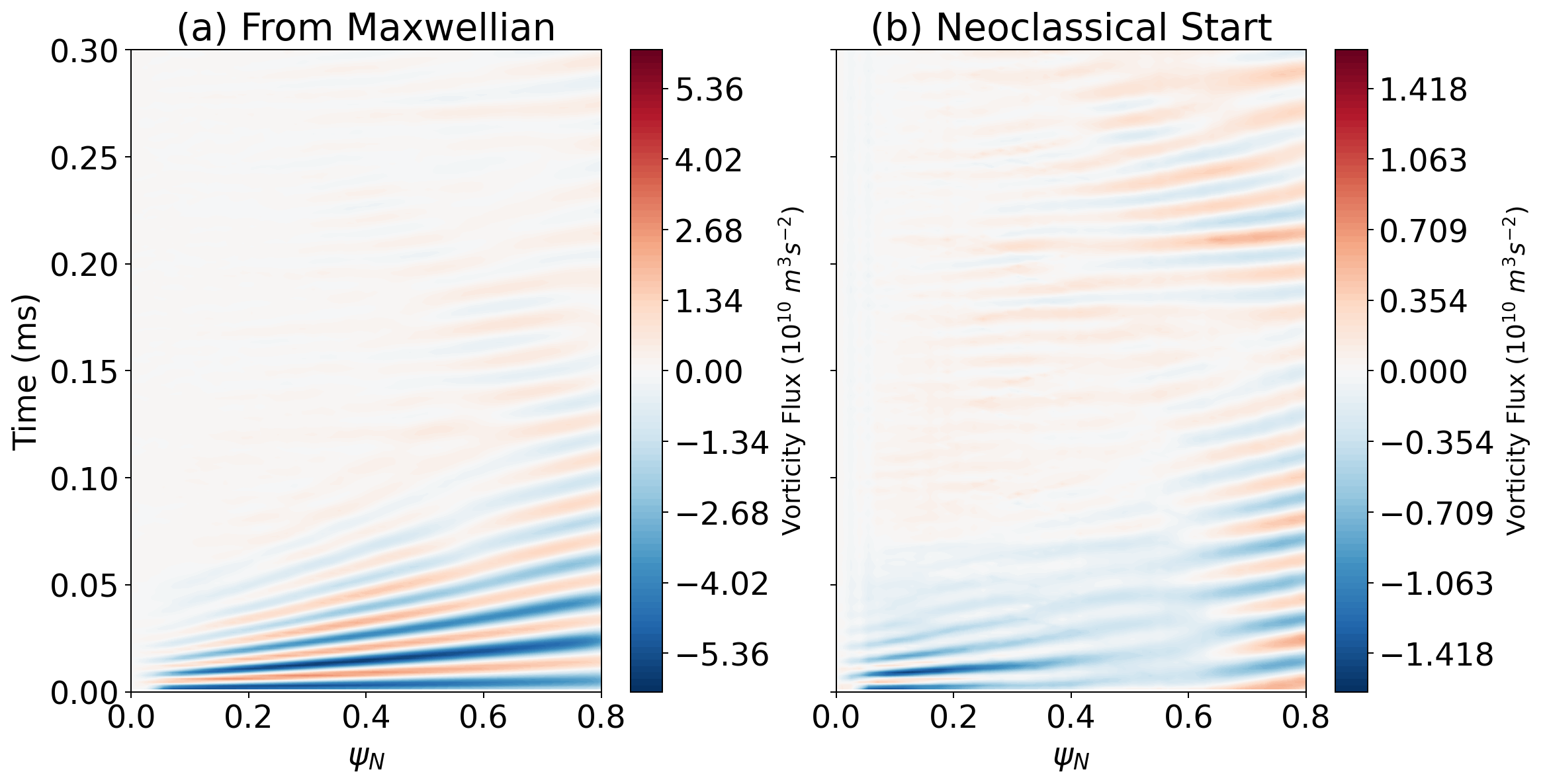}
\caption{Radial vorticity-flux transients in the axisymmetric CBC relaxation tests as a function of normalized poloidal flux $\psi_N$ and elapsed time since initialization. (a) Maxwellian initialization. (b) Neoclassical initialization using the distribution extracted at $t=0.31$ ms and smoothed with widths of $\Delta\theta=0.5$ rad in poloidal angle and $0.5\Delta v$ in velocity space. Each panel uses its own labeled diverging color scale centered at zero. The color range in (b) is approximately one quarter of that in (a), making the residual oscillations visible. Neoclassical initialization reduces the transient amplitude throughout the plotted radial interval.}
\label{fig:GAM2d}
\end{figure}

The radial structure of this reduction is shown in Fig.~\ref{fig:GAM2d}. With Maxwellian initialization, coherent GAM bands appear over the entire radial interval and persist during the early relaxation phase. In contrast, the neoclassical start with $\Delta\theta=0.5$ rad smoothing strongly suppresses these bands at all radii shown, leaving only a much weaker residual oscillation. This confirms that the reduction is a global suppression of the initial GAM response throughout the CBC radial domain.

The axisymmetric CBC tests therefore distinguish two benefits: the relaxed non-Maxwellian initial state substantially reduces the initial transient, while poloidal smoothing reduces the subsequent residual oscillations and marker-weight growth.

\begin{figure}
\includegraphics[width=0.58\textwidth]{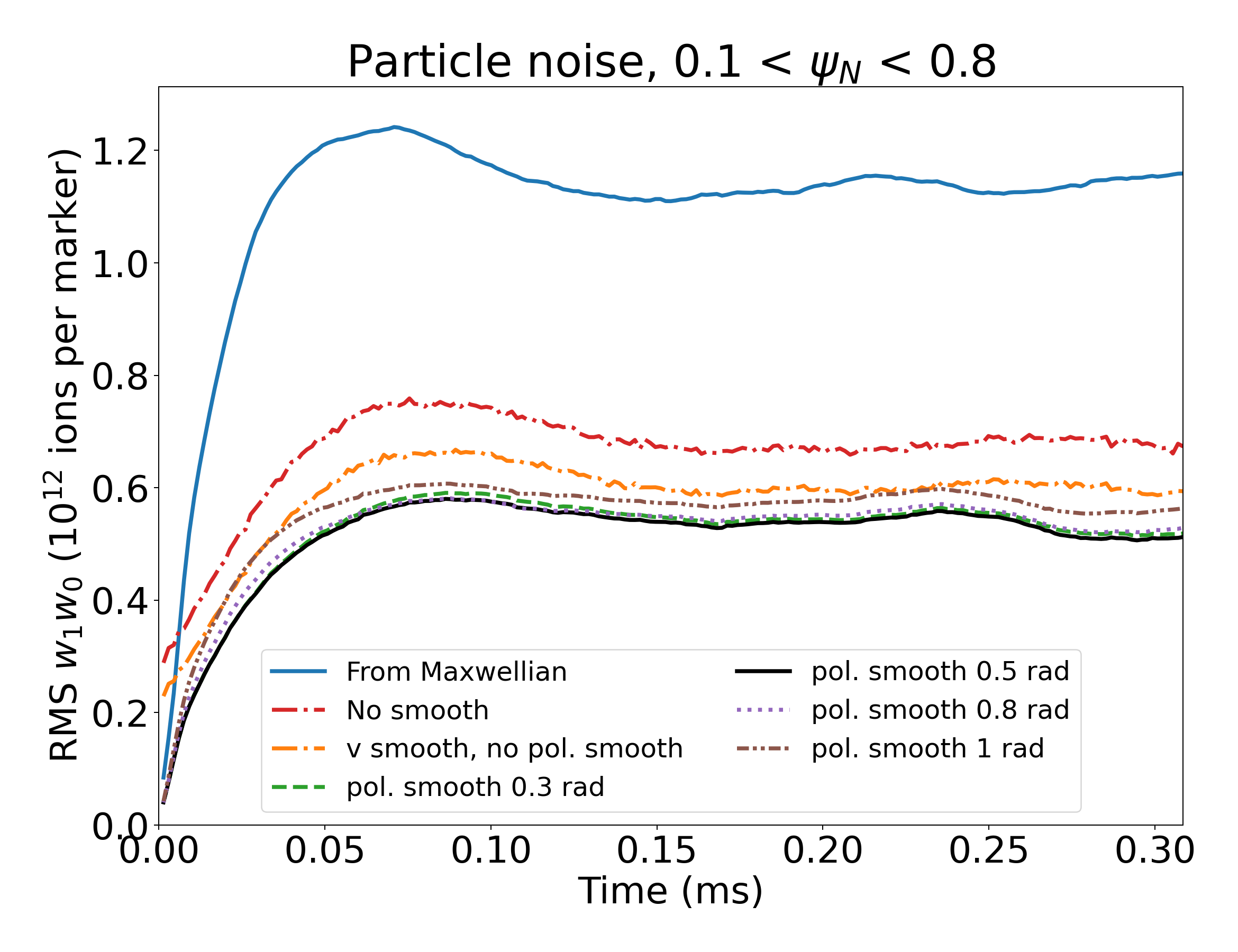}
\caption{Root-mean-square marker weights in the axisymmetric CBC relaxation tests, used as a proxy for particle noise. The quantity $W_{\rm rms}$ is defined in Eq.~(\ref{eq:wrms}) as the rms of $w_1w_0$ over sampled markers in $0.1<\psi_N<0.8$ and is plotted in units of $10^{12}$ physical ions per marker. The initialization methods and line styles are the same as in Fig.~\ref{fig:GAM}. Here, time is measured from the start of each run.}
\label{fig:cbc_noise}
\end{figure}

\begin{figure}
\includegraphics[width=0.58\textwidth]{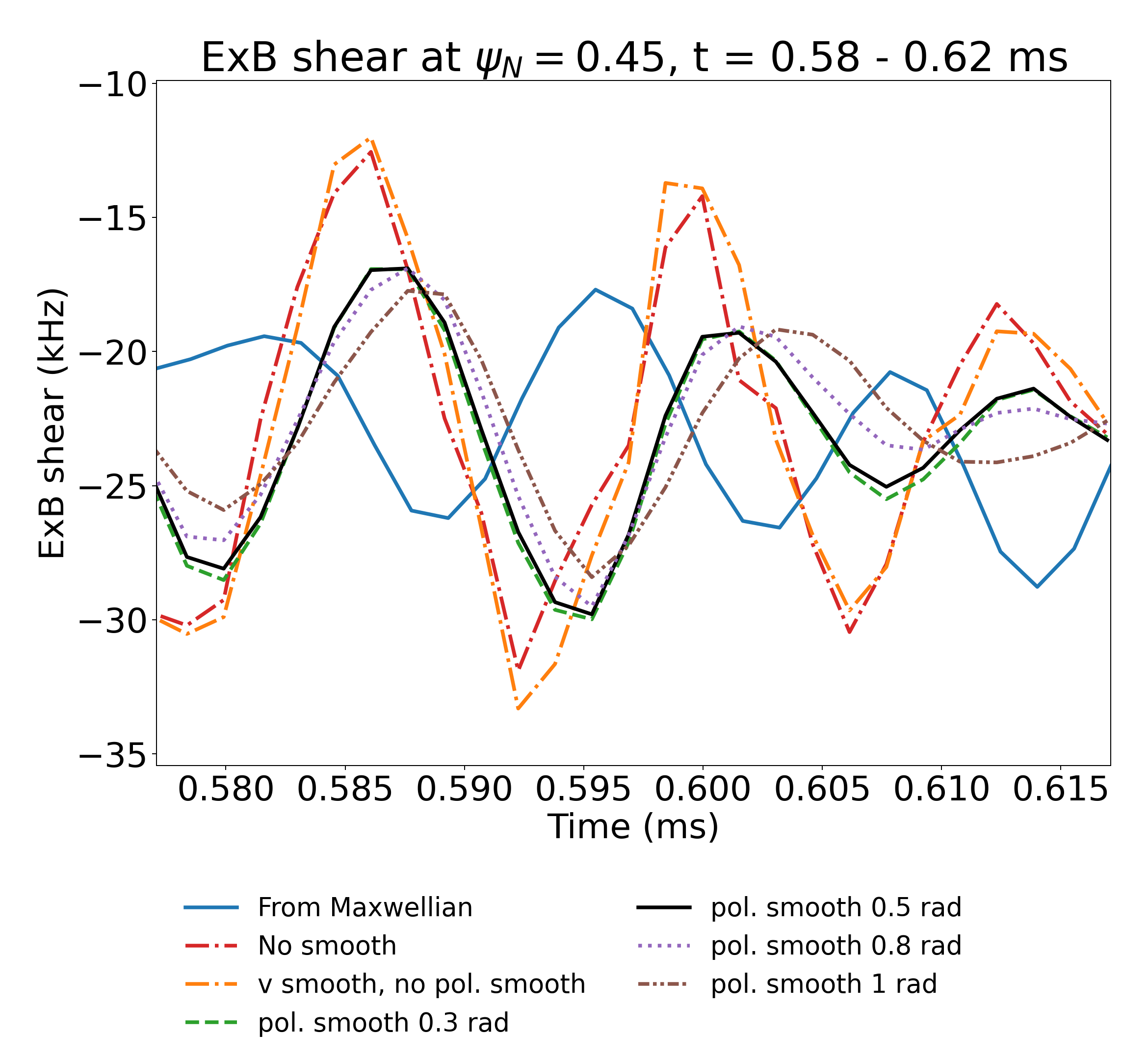}
\caption{Late-time $E\times B$ shear oscillations at $\psi_N=0.45$ in the axisymmetric CBC relaxation tests. This is an enlarged view of Fig.~\ref{fig:GAM}(b) over $t=0.58$--$0.62$ ms, corresponding to $0.27$--$0.31$ ms after neoclassical initialization. The initialization methods and line styles are the same as in Fig.~\ref{fig:GAM}.}
\label{fig:GAM_end}
\end{figure}

The smoothing-width scan also shows how distribution conditioning affects subsequent marker-weight growth. We use the root-mean-square marker weight associated with $f_{\rm ptl}$ as a proxy for particle noise,
\begin{equation}
W_{\rm rms}(t) = \left[ \frac{1}{N_m} \sum_{j=1}^{N_m} \left( w_{1,j}\, w_{0,j} \right)^2 \right]^{1/2},
\label{eq:wrms}
\end{equation}
where $w_{0,j}$ is the full-f weight of marker $j$, expressed as the number of physical ions it represents. The dimensionless weight $w_{1,j}$ specifies the fraction of this contribution associated with $f_{\rm ptl}=f-f_M-f_{\rm grid}$, so that $w_{1,j}w_{0,j}$ is the marker contribution to $f_{\rm ptl}$ in physical-ion units. The sum includes the $N_m$ markers in $0.1<\psi_N<0.8$ from the saved sample, which contains every 1000th marker. Thus, $W_{\rm rms}$ measures the rms magnitude of the correction represented by the markers and vanishes when the background $f_M+f_{\rm grid}$ equals the full distribution. Because the correction also contains physical changes of the distribution, a lower $W_{\rm rms}$ indicates a smaller marker correction; we use it as an indicator, not a direct measurement, of lower particle noise. As shown in Fig.~\ref{fig:cbc_noise}, neoclassical initialization substantially lowers $W_{\rm rms}$ after the initial adjustment. Among the tested widths, $\Delta\theta=0.3$--$0.5$ rad gives the lowest levels. The unsmoothed distribution and velocity-space smoothing alone produce higher levels; increasing the poloidal width to $0.8$--$1$ rad also raises $W_{\rm rms}$ slightly.
Figure~\ref{fig:GAM_end} shows a corresponding reduction in the late-time $E\times B$ shear oscillations. Without poloidal smoothing, their amplitude is comparable to that in the Maxwellian run over the same plotted time interval. With poloidal smoothing, the amplitude is approximately halved. This trend is consistent with the lower marker-weight levels in Fig.~\ref{fig:cbc_noise}.
For this CBC scan, $\Delta\theta=0.5$ rad provides a useful balance: it retains an initial transient comparable to that of the unsmoothed neoclassical start, reduces the late-time oscillations, and yields one of the lowest marker-weight levels. The optimal width may differ for other equilibria and for a different number of markers per velocity-space cell. In general, the smoothing width should be chosen to balance two requirements: the distribution must be smooth enough to minimize the effect of marker noise, yet the smoothing must not significantly perturb the dominant low-$m$ neoclassical poloidal structure.

Quantitatively, neoclassical initialization with $\Delta\theta=0.5$ rad smoothing reduces the rms marker weight by about a factor of two relative to Maxwellian initialization. Under the usual $1/\sqrt{N_p}$ scaling of sampling noise, with otherwise comparable marker statistics and normalization, this reduction suggests a noise level equivalent to that obtained with roughly four times as many markers. This is a scaling estimate; the comparisons presented here do not demonstrate that the marker population can be reduced to one quarter while preserving the turbulent dynamics. Such a reduction would require a separate marker-number convergence study. The observed weight reduction is consistent with a background $f_M+f_{\rm grid}$ close to the neoclassical pseudo-equilibrium, leaving a smaller correction $f_{\rm ptl}$ to be represented by the markers.

\subsection{Turbulence simulation}\label{sec:cbc_turb}

After interpolation of the conditioned distribution to the turbulence mesh, the CBC turbulence simulation shows the same qualitative behavior as the axisymmetric precursor tests. The initial GAM transient remains substantially smaller than in the Maxwellian-initialized case, and the particle-weight level is reduced by more than a factor of two. The reduction in weight variance is most important during the early phase of the turbulence evolution, when the instability seed level and the clean onset of the linear growth stage are most sensitive to initialization noise.

\begin{figure}
\includegraphics[width=0.68\textwidth]{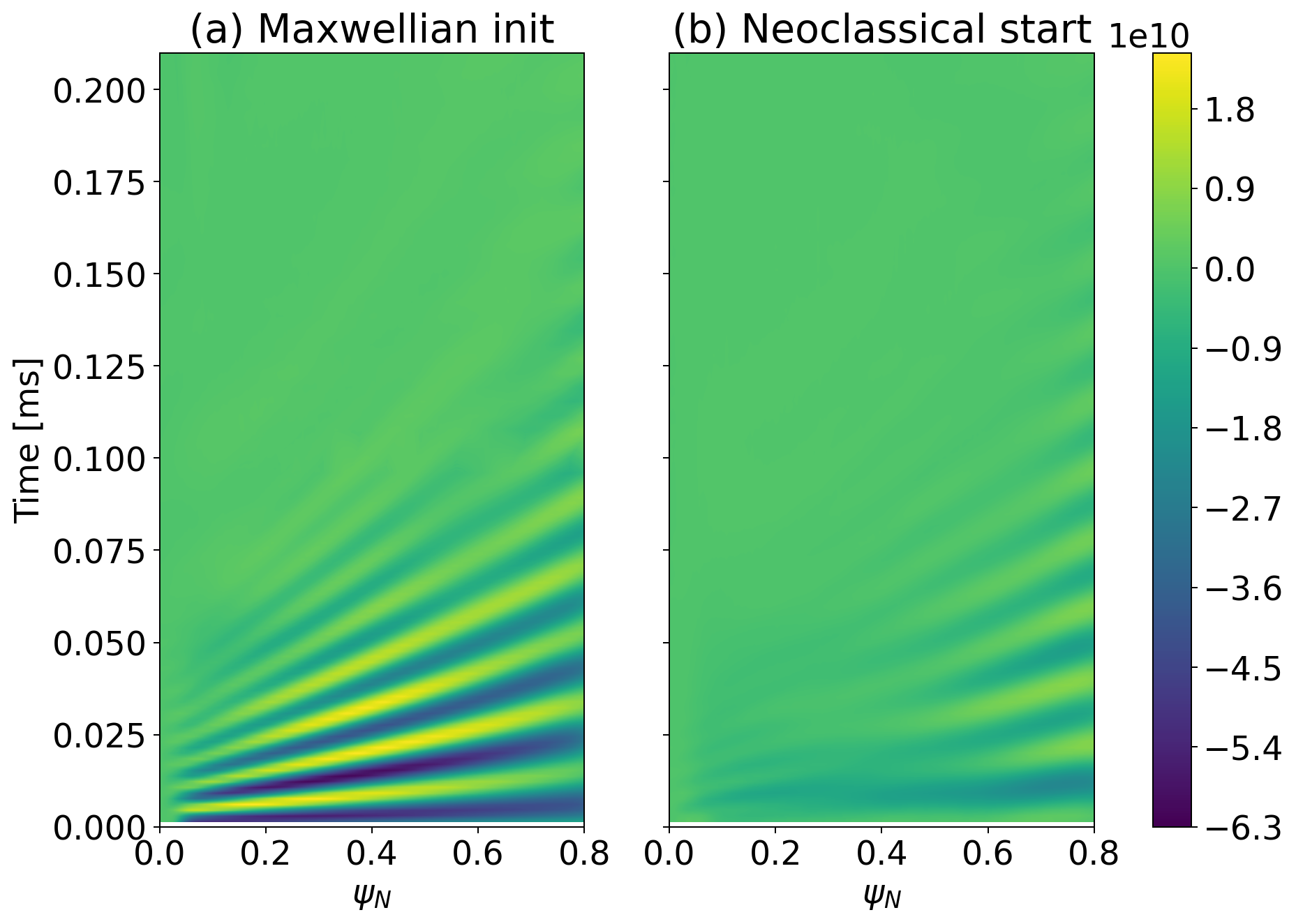}
\caption{Initial transient oscillations of the radial vorticity flux in the CBC turbulence simulation as a function of normalized poloidal flux $\psi_N$ and time. (a) Maxwellian initialization. (b) Neoclassical initialization using the conditioned distribution from the axisymmetric precursor calculation. The same color scale is used for both panels, showing that the neoclassical start substantially suppresses the initial GAM response after transfer to the turbulence mesh.}
\label{fig:GAM2d_turb}
\end{figure}

\begin{figure}
\includegraphics[width=0.68\textwidth]{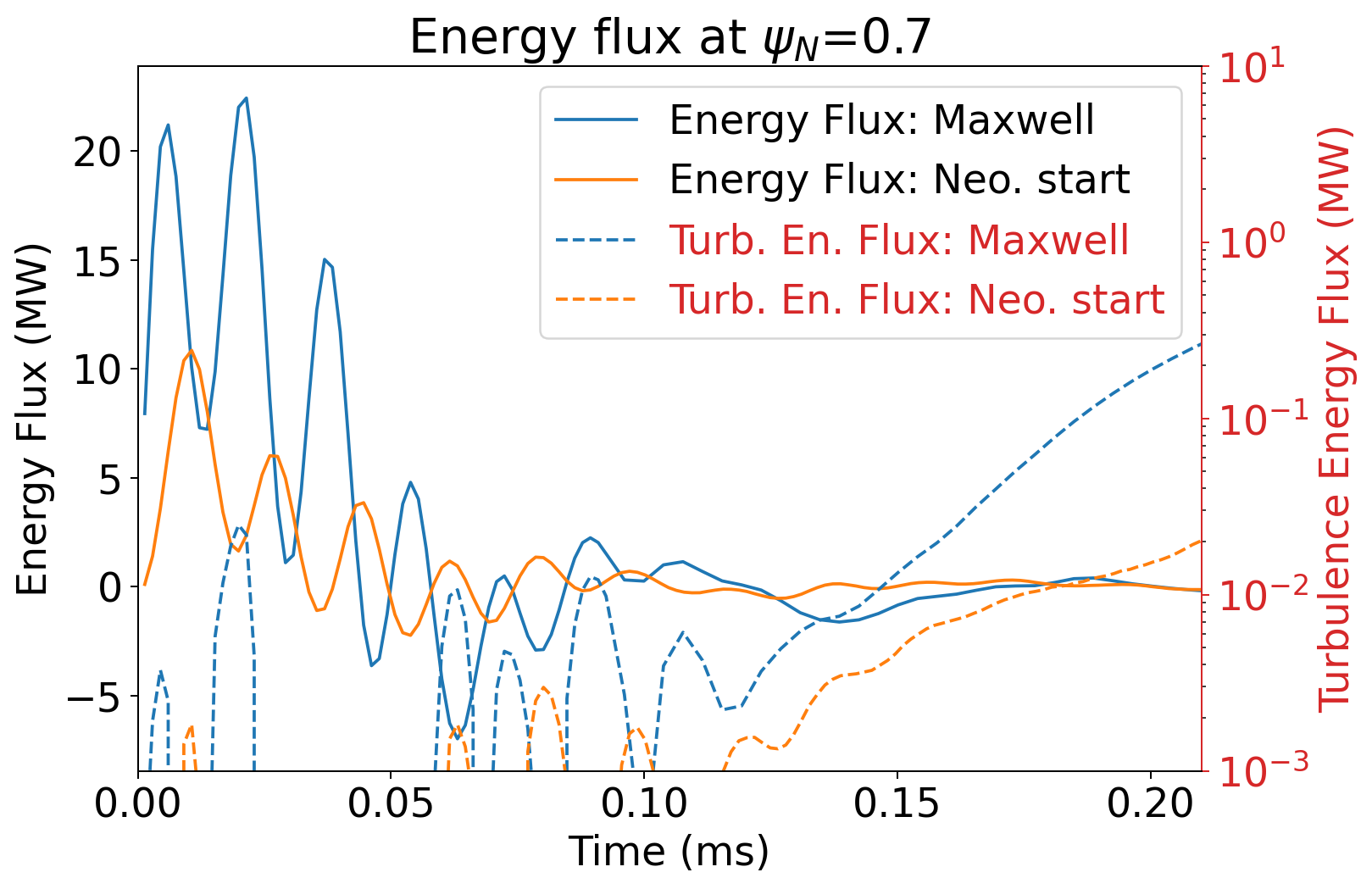}
\caption{The radial energy flux in the CBC turbulence simulation at $\psi_N=0.7$. Blue lines represent Maxwellian initialization and orange lines represent neoclassical initialization. 
Solid lines show the total radial ion energy flux $Q_i$ of Eq.~(\ref{eq:eflux}), including the transient GAM contribution, and correspond to the left vertical axis.
Dashed lines show its $E\times B$ part $Q_i^{E}$, i.e. the turbulence transport, and correspond to the logarithmic right vertical axis; gaps in the dashed lines are times at which $Q_i^{E}$ is negative, which occur only during the initial transient, while $Q_i^E$ is still small. Positive values of the fluxes are outward.}
\label{fig:GAM_turb}
\end{figure}

Fig.~\ref{fig:GAM2d_turb} confirms that the reduction of the initial GAM response survives the transfer from the axisymmetric precursor calculation to the turbulence simulation. With Maxwellian initialization, coherent oscillatory bands appear over the radial domain during the early phase of the calculation, similar to the axisymmetric relaxation test. In contrast, the neoclassical start strongly reduces the radial vorticity-flux oscillation throughout the domain, leaving only a weak residual response. This demonstrates that the conditioned neoclassical distribution and its associated fields remain sufficiently close to the relaxed state after interpolation to the turbulence mesh.

Fig.~\ref{fig:GAM_turb} shows the radial energy flux including the effect of the initial transient GAM (left vertical axis) and the radial turbulence energy flux with exponential growth (right vertical axis with log scale). The $E\times B$ part $Q_i^E$ is used as the turbulent flux because the GAM response is carried mainly by the radial magnetic-drift ($\nabla B$ and curvature) orbit motion of the ions, so that the radial $E\times B$ drift, which excludes the magnetic drift, transports energy mainly through turbulence rather than through the GAM. 
In the axisymmetric CBC test with Maxwellian initialization, where no turbulence is present, the maximum of $|Q_i^E|$ over the whole run ($0\le t\le0.62$ ms) is below $0.4\%$ of the maximum of $|Q_i|$ at both $\psi_N=0.45$ and $0.7$. The total flux is negative at some times before $t\simeq0.15$ ms, which with the present sign convention is an inward energy flux; it arises from the GAM oscillation, which alternates in sign. The turbulence activity is higher with Maxwellian initialization since 
the higher particle-noise level provides a larger initial seed for the turbulence.
Note that the simulation is inherently nonlinear due to the nonlinear interaction between the unstable turbulence modes and the transient GAM oscillations. 
In the CBC turbulence results, the benefit is strongest during the "linear-growth" interval or "pre-linear-growth" interval, approximately $t < 0.1$ ms. At later times, after the transient has decayed and the turbulence approaches the nonlinear state, the difference between the two initialization methods becomes less pronounced. 
This behavior is physically reasonable: once the system has evolved away from its initial condition and the turbulence self-organizes, the memory of the initialization is reduced. The main advantage of the neoclassical initialization is therefore not to alter the eventual nonlinear state in a dramatic way, but to provide a cleaner and less contaminated path into that state. Dif-Pradalier \textit{et al.}\cite{Dif-Pradalier_2008} showed that full-f simulations started from a local Maxwellian and from a canonical Maxwellian reach turbulent states with nearly identical statistical properties after a long transient. In this study, the convergence to the same saturated state has not been verified because of the short simulations. We nevertheless expect it, since the two initial states compared here are much closer to each other than those in Ref.~\onlinecite{Dif-Pradalier_2008}.

\section{ASDEX Upgrade I-mode Demonstration\label{sec:asdex}}

As a second demonstration, we apply the initialization scheme to a realistic diverted tokamak geometry based on ASDEX Upgrade I-mode discharge \#37980 at $t=4.2$ s \cite{Pitzal2026}. This case is more demanding than the CBC test because the pedestal and scrape-off-layer regions introduce strong radial variation in the turbulence dynamics and in the zonal-flow response. In particular, edge turbulence develops faster than core turbulence, so long-lived initialization-driven GAMs can overlap directly with the time interval used for pedestal transport analysis. The I-mode scenario was chosen because the interaction between GAMs and edge turbulence in I-mode has been observed experimentally\cite{Manz_2015}; simulation studies of this interaction require the initialization-driven GAM to be isolated from the physical one.

\begin{figure}
\includegraphics[width=0.58\textwidth]{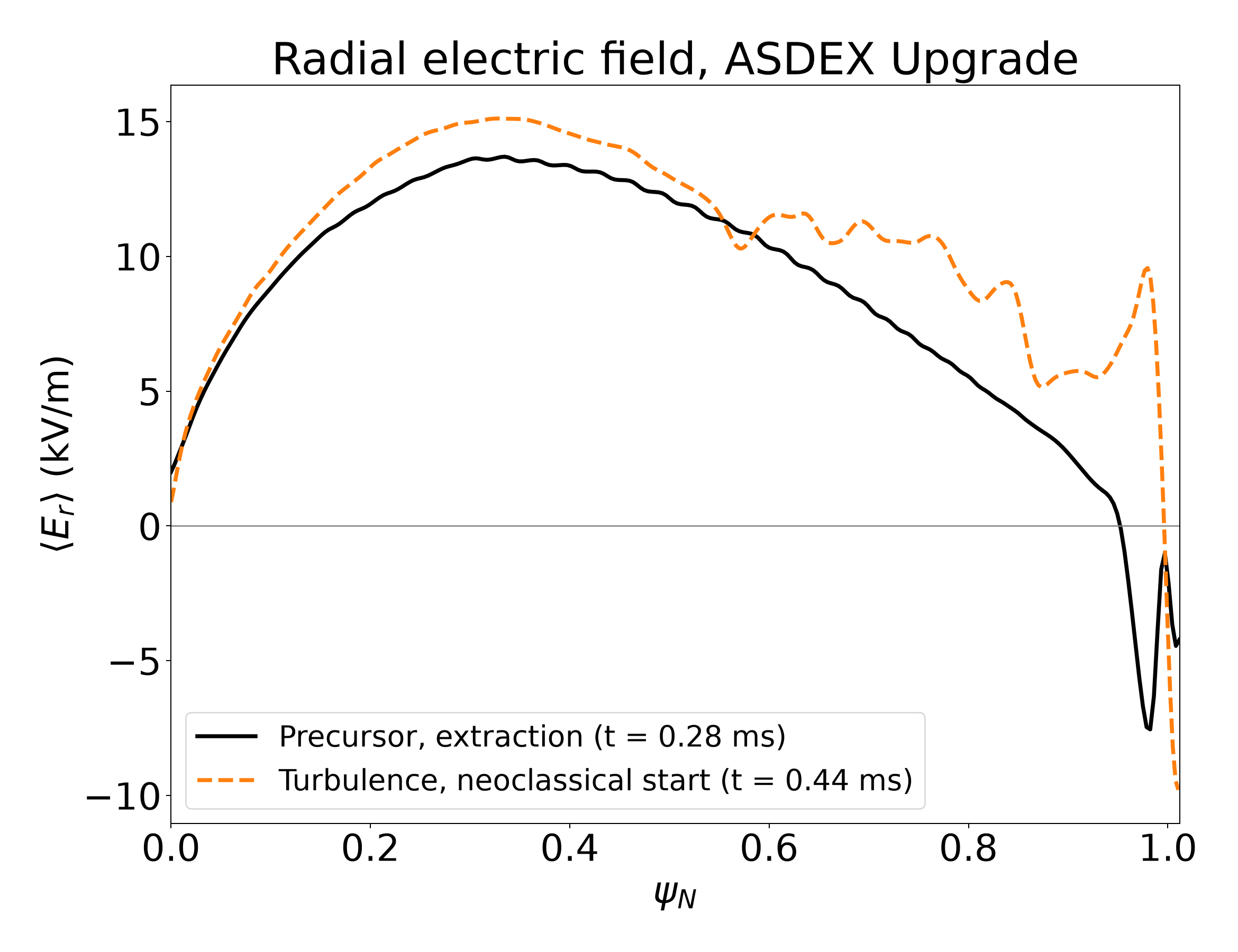}
\caption{Flux-surface-averaged radial electric field $\langle E_r\rangle$, defined as in Fig.~\ref{fig:er_cbc}, for ASDEX Upgrade: the axisymmetric precursor at the distribution extraction time, $t=0.28$ ms (solid), and the turbulence simulation started from the neoclassical initial state at the end of the run, $t=0.44$ ms (dashed). Both profiles are smoothed with a Gaussian of width $0.005$ in $\psi_N$, and the plotted interval ends at $\psi_N=1.012$.}
\label{fig:er_aug}
\end{figure}

\begin{figure}
\includegraphics[width=0.38\textwidth]{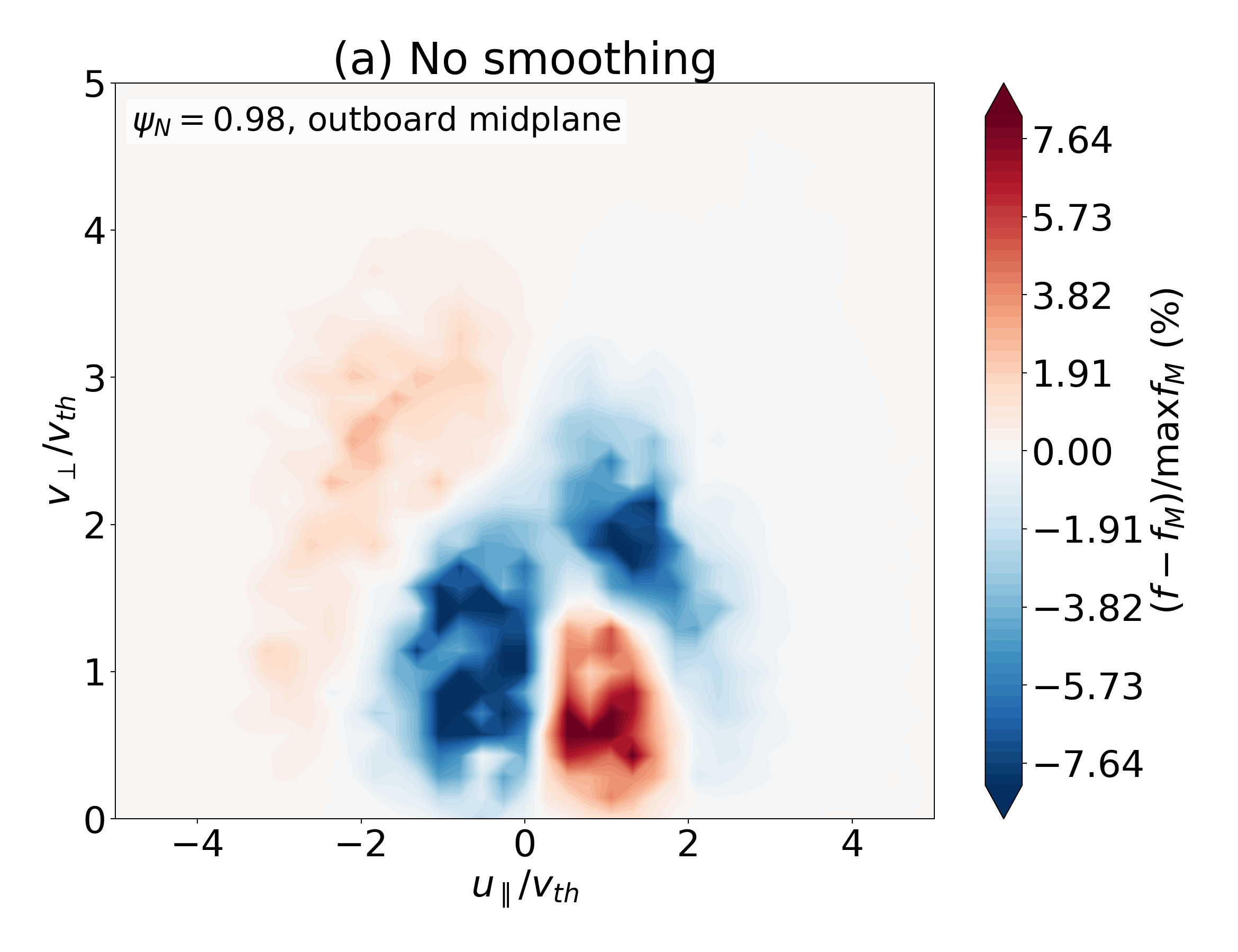}
\includegraphics[width=0.38\textwidth]{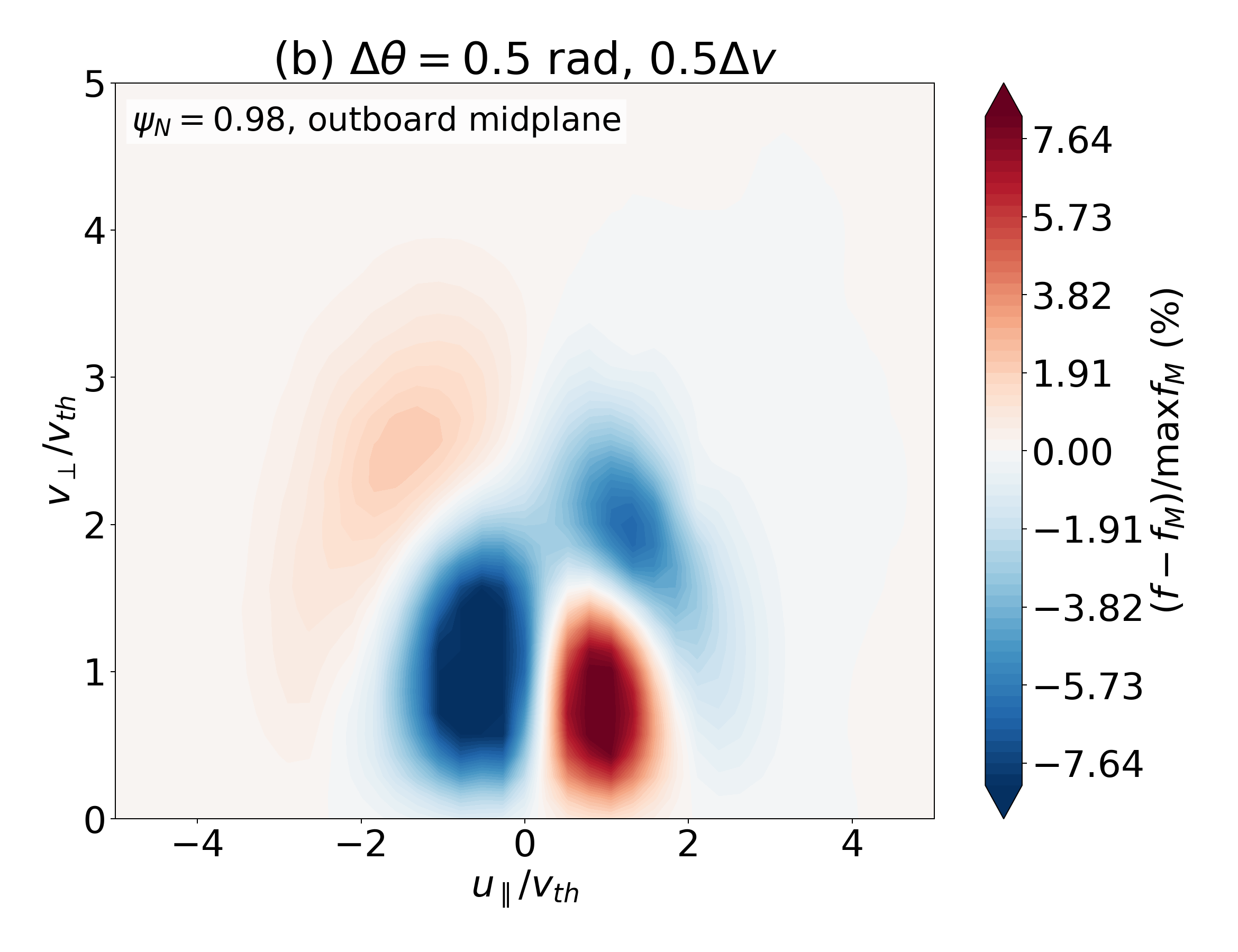}
\caption{Deviation of the extracted ASDEX Upgrade distribution from the local Maxwellian, $100(f-f_M)/\max f_M$ (\%), in the same format as Fig.~\ref{fig:distribution}. The sample is taken at the outboard midplane, $\psi_N=0.98$ ($R=2.126$ m, $Z=0.009$ m), in the axisymmetric precursor at $t=0.28$ ms. (a) Without smoothing. (b) After Gaussian smoothing with widths of $0.5$ rad in poloidal angle and $0.5\Delta v$ in velocity space. The color range is clipped at the 98th percentile of panel (a).}
\label{fig:distribution_aug}
\end{figure}

The axisymmetric precursor for this case starts from a local Maxwellian, and the distribution is extracted at $t=0.28$ ms, after the initialization-driven GAM has largely decayed. Only the ion distribution is transferred; in the turbulence simulation started from the neoclassical state, the electrons are initialized with a local Maxwellian. Figure~\ref{fig:er_aug} compares the radial electric field of the precursor at the extraction time with that of the turbulence simulation started from the neoclassical initial state at the end of the run, $t=0.44$ ms. In the precursor, $\langle E_r\rangle$ is positive over most of the core, in contrast to the CBC case, reaching about $14$ kV/m near $\psi_N\simeq0.33$, and forms a negative well of about $-7.5$ kV/m at $\psi_N\simeq0.98$ inside the separatrix. For $\psi_N\lesssim0.55$ the turbulence simulation retains this profile to within about $10\%$. Farther out its field is larger, and near the pedestal the well is replaced by a positive peak of about $10$ kV/m at $\psi_N\simeq0.98$, with the field turning negative only at the separatrix. This edge difference reflects the evolution of the pedestal during the turbulence simulation, which includes heating. Figure~\ref{fig:distribution_aug} shows the extracted distribution at $\psi_N=0.98$ on the outboard midplane. The deviation from the local Maxwellian reaches $16\%$ of $\max f_M$ before smoothing ($2.2\%$ in rms over velocity space) and $12.5\%$ after smoothing ($2.2\%$ in rms), so the non-Maxwellian structure carried into the turbulence calculation is larger in the pedestal than in the CBC core.

\begin{figure}
\includegraphics[width=0.72\textwidth]{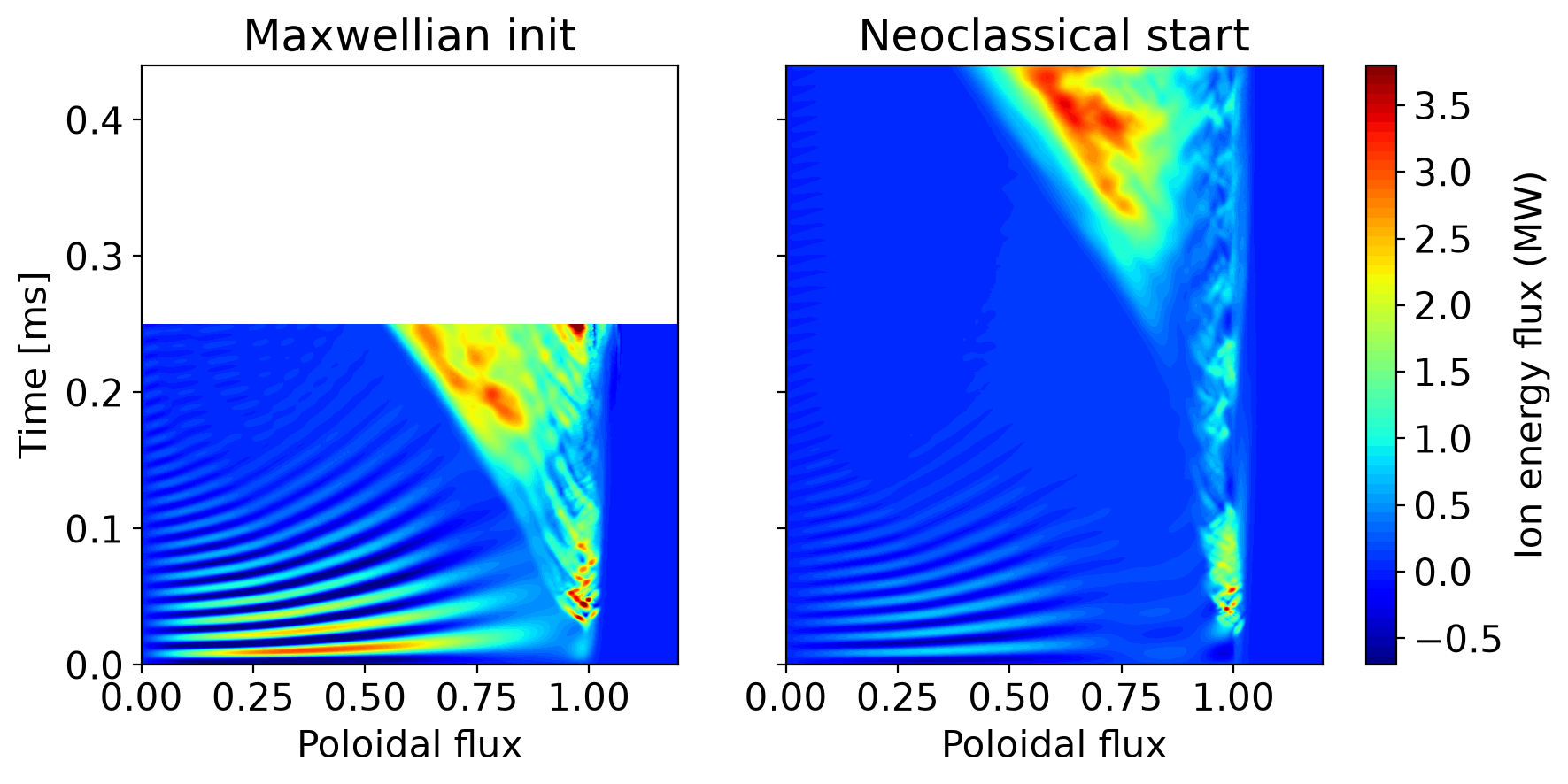}
\caption{Radial ion energy flux $Q_i$ [Eq.~(\ref{eq:eflux})] in the ASDEX Upgrade I-mode turbulence simulation. The Maxwellian initialization produces GAM oscillations associated with the initial transient, which persist into the early turbulence phase. The neoclassical start substantially reduces this initial oscillatory component, allowing the turbulence-driven energy flux to emerge with less contamination from the initialization transient.}
\label{fig:asdex_turb_en_flux}
\end{figure}

Fig.~\ref{fig:asdex_turb_en_flux} compares the radial ion energy-flux evolution obtained from Maxwellian and neoclassical initialization. The simulation with Maxwellian initialization (left) is shown up to $t=0.25$ ms; a numerical instability developed after this time, and the simulation was stopped. The simulation with neoclassical initialization (right) continued further to observe core turbulence development. In the Maxwellian-initialized case, coherent oscillatory bands are visible over a broad radial region during the early phase of the simulation. These bands are the transport signature of the initialization-driven GAM and persist even as edge turbulence begins to develop. As a result, the measured radial ion energy flux contains a mixture of physical turbulent transport and artificial axisymmetric relaxation.

The neoclassical start strongly reduces this coherent GAM component, making the turbulence-driven structures easier to distinguish in the energy-flux evolution. This is particularly useful near the pedestal top, where GAM-induced $E\times B$ shear can interfere with turbulence onset and with the inferred transport level. Suppressing the initialization transient therefore improves the interpretability of early-time pedestal transport.

For the ASDEX Upgrade case, the axisymmetric precursor evolves only the $n=0$ dynamics on a coarser mesh than the turbulence calculation. The precursor used 8 GPU nodes on NERSC Perlmutter for 4 h, about $32$ node-hours, to reach $t=0.28$ ms. The production turbulence simulation used 128 GPU nodes for about $155$ h of wall-clock time in total, including restarts, or about $2\times10^4$ node-hours, to cover about $0.44$ ms; reaching the same simulated time as the precursor, $0.28$ ms, took about $1.3\times10^4$ node-hours. The precursor was therefore about $400$ times cheaper for the same simulated time, and the relaxed initial state was obtained at a small additional cost relative to the production run.

In the axisymmetric precursor calculation for this case, which started from a local Maxwellian, the initialization-driven GAM required approximately $0.3$ ms to decay. This decay time illustrates the persistence of the initial transient, but it does not measure the production runtime that can be saved by suppressing it. The present comparisons demonstrate reduced initialization transients and, in the CBC tests, lower marker weights, indicating reduced particle noise, over the simulated interval. They do not establish a reduction in the time or computational cost required to obtain statistically converged turbulent transport. Quantifying such a reduction would require both initializations to satisfy the same convergence criterion and would need to include the precursor and distribution-transfer costs.

Together, the CBC and ASDEX Upgrade results establish two main benefits: lower marker weights, indicating reduced particle noise and suppression of initialization-driven transients in both cases. These improvements provide a cleaner initial state and reduce contamination of the early turbulence and transport signals.

\section{Conclusion\label{sec:conclusion}}
We have developed a neoclassical initialization method for total-f gyrokinetic turbulence simulations that directly addresses the artificial transient GAMs generated by local-Maxwellian loading. Instead of initializing the multiscale multiphysics simulation with an analytic Maxwellian equilibrium, the method uses an inexpensive axisymmetric preconditioning calculation to obtain a numerically relaxed non-Maxwellian axisymmetric distribution together with its self-consistent radial electric field. After phase-space smoothing and interpolation to the target mesh, this relaxed state is used as the initial condition for the total-f production simulation.

The Cyclone Base Case tests show that the relaxed initial state substantially suppresses the initialization-driven response throughout the plotted radial domain. At $\psi_N=0.45$, the peak absolute vorticity flux decreases from about $6$ to $0.3$--$1.3$ in the plotted units, while the peak absolute $E\times B$ shear decreases from about $140$ to $30$--$40$ kHz. The smoothing scan separates this transient reduction from the benefits of poloidal filtering. Widths of $\Delta\theta=0.3$--$0.5$ rad retain an initial transient comparable to that of the unsmoothed neoclassical start, approximately halve the late-time shear oscillations, and yield the lowest rms marker weights among the tested cases. Wider filters of $0.8$--$1$ rad produce larger initial vorticity-flux excursions, consistent with distortion of the relaxed state. The tested width of $0.5$ rad is about $8\%$ of the $2\pi$ angular wavelength of an $m=1$ structure, but this scale comparison does not establish a universal optimum. The factor-of-two reduction in rms marker weight suggests a fourfold equivalent marker population under standard sampling-noise scaling; a corresponding reduction in the required marker count has not been demonstrated.

The ASDEX Upgrade I-mode demonstration confirms that the method remains effective in realistic diverted-edge geometry. In this case, Maxwellian initialization produces GAM oscillations in the radial ion energy flux that persist into the early turbulence phase and can interfere with pedestal transport analysis. Neoclassical initialization strongly reduces this GAM component, allowing the turbulence-driven transport signal to emerge more cleanly. This result extends the demonstrated suppression of initialization transients from the CBC test to a realistic edge configuration.

The principal advantages of the method are lower marker weights, indicating reduced particle noise, and suppression of initial transients. Starting from a numerically relaxed distribution reduces the marker-weight correction and the artificial axisymmetric response associated with local-Maxwellian loading. The resulting initial state provides a cleaner basis for studying the early evolution of turbulence and interpreting transport in total-f gyrokinetic simulations. Although the demonstrations shown here are electrostatic, the scheme only prepares the initial state and does not change the evolution equations of the production run, apart from the interpolation error of the transferred state; it is therefore equally applicable to electromagnetic simulations. The flux-surface averages it uses are defined on the equilibrium flux surfaces before turbulent perturbations develop.

\begin{acknowledgments}

This research was supported by the U.S. Department of Energy, Office of Science, Offices of Fusion Energy Sciences (FES) and Advanced Scientific Computing Research (ASCR), through the SciDAC-5 Partnership Center, Computational Evaluation and Design of Actuators for Core-Edge Integration (CEDA), under Award Number DE-AC02-09CH11466.

This research used resources of the National Energy Research Scientific Computing Center (NERSC), a Department of Energy Office of Science User Facility using NERSC award FES-ERCAP0036720 and FES-ERCAP0037492.

An award of computer time was provided by the INCITE program. This research also used resources of the Oak Ridge Leadership Computing Facility, which is a DOE Office of Science User Facility supported under Contract DE-AC05-00OR22725.

\end{acknowledgments}

\section*{Data Availability Statement}
Raw data are stored at the National Energy Research Scientific Computing Center (NERSC). Derived data supporting the findings of this study are available from the corresponding author upon reasonable request.

\section*{Disclaimer}
This report was prepared as an account of work sponsored by an agency of the United States Government. Neither the United States Government nor any agency thereof, nor any of their employees, makes any warranty, express or implied, or assumes any legal liability or responsibility for the accuracy, completeness, or usefulness of any information, apparatus, product, or process disclosed, or represents that its use would not infringe privately owned rights. Reference herein to any specific commercial product, process, or service by trade name, trademark, manufacturer, or otherwise does not necessarily constitute or imply its endorsement, recommendation, or favoring by the United States Government or any agency thereof. The views and opinions of authors expressed herein do not necessarily state or reflect those of the United States Government or any agency thereof.

\appendix

\section{Identification of the transient as a GAM}\label{app:gam}

To identify the axisymmetric transient in the CBC tests, we Fourier transformed the flux-surface-averaged $E\times B$ shear $\omega_E(\psi,t)$ of Eq.~(\ref{eq:shear}) in time at each flux surface for the Maxwellian-initialized precursor run. The transform covers $0\le t\le0.31$ ms, the interval before the distribution is extracted. The one-dimensional diagnostic is written every $1.54~\mu$s, and the window gives a frequency resolution of $3.2$ kHz. Before the transform, a slow drift is removed by subtracting a Gaussian running mean of width $25~\mu$s, which leaves the GAM frequency range essentially unaffected, and a Hann window is applied.

\begin{figure}
\includegraphics[width=0.55\textwidth]{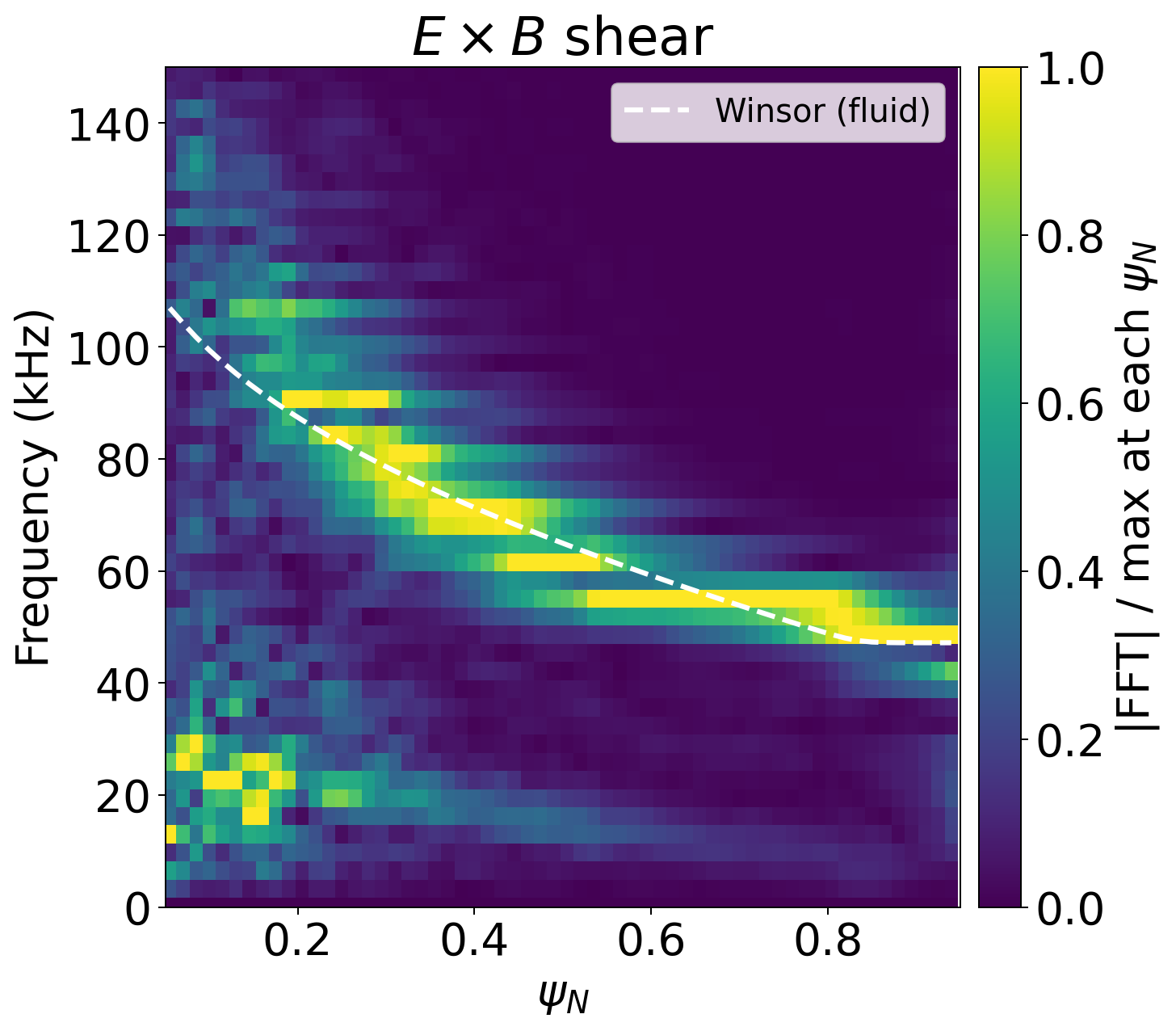}
\caption{Frequency spectrum of the flux-surface-averaged $E\times B$ shear in the Maxwellian-initialized axisymmetric CBC run over $0\le t\le0.31$ ms, as a function of $\psi_N$. The amplitude is normalized to its maximum at each $\psi_N$. The dashed line is the fluid GAM frequency of Eq.~(\ref{eq:fgam}).}
\label{fig:gam_spectrum}
\end{figure}

Figure~\ref{fig:gam_spectrum} shows the resulting amplitude, normalized to its maximum at each $\psi_N$, together with the fluid GAM frequency \cite{Winsor_1968}
\beq
f_{\rm GAM} = \frac{1}{2\pi R_0}\left[\left(\frac{7}{4}+\frac{T_e}{T_i}\right)\frac{2T_i}{m_i}\right]^{1/2},
\label{eq:fgam}
\eeq
evaluated with the initial ion temperature profile, $T_e/T_i=1$, and the major radius of the magnetic axis, $R_0=1.70$ m. The spectral peak follows $f_{\rm GAM}$ from about $90$ kHz at $\psi_N\simeq0.2$ to about $50$ kHz at $\psi_N\simeq0.8$. Over $0.2\le\psi_N\le0.85$, the median ratio of the peak frequency to $f_{\rm GAM}$ is $1.01$. Near the magnetic axis, $\psi_N\lesssim0.2$, the spectrum is broad and no single peak is resolved. The initial transient oscillation is therefore the GAM.

\section{Distortion of the poloidal structure by smoothing}\label{app:distortion}

\begin{figure}
\includegraphics[width=0.75\textwidth]{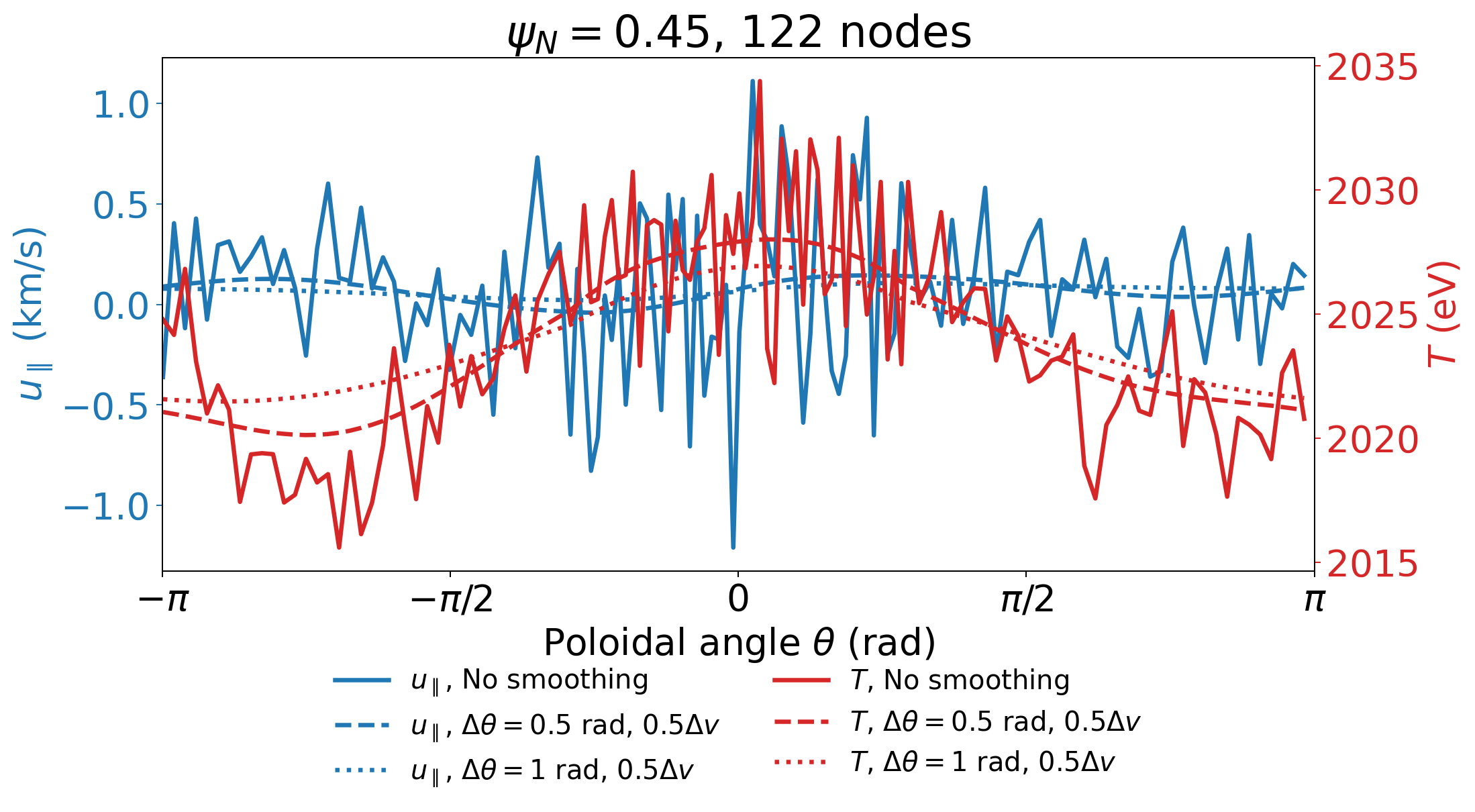}
\caption{Local parallel flow $u_\parallel$ (blue, left axis) and temperature $T$ (red, right axis) of the extracted CBC distribution on the flux surface $\psi_N=0.45$ as functions of the geometric poloidal angle $\theta$, with $\theta=0$ at the outboard midplane. Solid lines: no smoothing. Dashed and dotted lines: poloidal smoothing with $\Delta\theta=0.5$ and $1$ rad, respectively, each combined with velocity-space smoothing of $0.5\Delta v$.}
\label{fig:poloidal_moments}
\end{figure}

To quantify how much the poloidal filter alters the neoclassical structure, the local parallel flow $u_\parallel$ and temperature $T$ of the extracted CBC distribution, $f_M+f_{\rm grid}$, were computed node by node from its velocity moments on the flux surface $\psi_N=0.45$, which has $122$ mesh nodes. Figure~\ref{fig:poloidal_moments} shows them as functions of the geometric poloidal angle $\theta$, measured from the outboard midplane, for the unsmoothed distribution and for poloidal smoothing widths of $\Delta\theta=0.5$ and $1$ rad, both combined with velocity-space smoothing of $0.5\Delta v$. The low-$m$ content is measured by Fourier coefficients in $\theta$.

The temperature has a clear $m=1$ variation of $4.4$ eV, $0.22\%$ of the poloidal-angle mean of $2024$ eV, with its maximum on the outboard side. Smoothing with $\Delta\theta=0.5$ rad reduces this amplitude by $12\%$, and smoothing with $\Delta\theta=1$ rad reduces it by $37\%$, close to the attenuation expected for a Gaussian filter in a uniform poloidal angle, $\exp(-m^2\Delta\theta^2/2)$, which gives $12\%$ and $39\%$ for $m=1$. The poloidal-angle mean is unchanged to within $0.02\%$. The parallel flow on this surface is small, with a mean of $0.07$ km/s, and before smoothing its poloidal variation is dominated by node-to-node fluctuations, with an unweighted standard deviation over the nodes of $0.38$ km/s. Smoothing reduces this node-to-node variation by $85\%$ at $\Delta\theta=0.5$ rad and by $93\%$ at $1$ rad; the remaining $m=1$ flow amplitude, $0.03$--$0.04$ km/s, is comparable to the noise level of the unsmoothed data. Thus $\Delta\theta=0.5$ rad removes most of the node-to-node variation while reducing the dominant $m=1$ structure by about one eighth, whereas $\Delta\theta=1$ rad removes more than one third of it. This is consistent with the larger initial transients of the starts with wider filters in Fig.~\ref{fig:GAM}. These values apply to this CBC flux surface. The kernel acts on the mesh-node index, the node spacing on this surface varies by a factor of about $1.8$, and velocity-space smoothing is applied as well, so the uniform-angle expression serves only as a guide; for a new case, the distortion should be checked on the actual mesh and moments.

\bibliography{references}

\end{document}